\documentclass[aps,jcp,superscriptaddress,amsmath,amssymb]{revtex4-1}

\usepackage{comment}
\usepackage{tikz}
\usetikzlibrary{trees}
\usetikzlibrary{decorations.pathmorphing}
\usetikzlibrary{decorations.markings}
\tikzset{
    photon/.style={decorate, decoration={snake,segment length=1.5mm}, draw=black},
    coulomb/.style={dotted},
    electron/.style={draw=black, postaction={decorate},
        decoration={markings,mark=at position .55 with {\arrow[draw=black]{>}}}}, 
    gluon/.style={decorate, draw=magenta,
        decoration={coil,amplitude=4pt, segment length=5pt}},
    boundelectron/.style={thick, double},
    transverse/.style={dashed}
}

\usepackage{graphicx}% Include figure files
\usepackage{dcolumn}% Align table columns on decimal point
\usepackage{bm}% bold math
\usepackage{rotating}

\newcolumntype{.}{D{.}{.}{8}}

\usepackage[utf8]{inputenc}
\usepackage{physics}
\usepackage{amsmath, amssymb}
\usepackage{tabularx,booktabs,array,dcolumn}
\usepackage{setspace}
\usepackage{vmargin}
\usepackage{xcolor}
\usepackage{graphicx,psfrag,subfigure}

\usepackage{float}
\usepackage[all]{xy}

\usepackage{bm}
\usepackage{mathtools}
\usepackage{physics}
\usepackage[english]{babel}

\newcommand{\bos}[1]{\boldsymbol{#1}}
\newcommand{\mr}[1]{\mathrm{#1}}
\newcommand{\pd}[2]{\frac{\partial #1}{\partial #2}}

\def\Eh{E_\mathrm{h}}

\def\iim{\mr{i}}

\def\DC{\text{DC}}
\def\Breit{\text{B}}
\def\DCB{\text{DCB}}
\def\DCpB{{\text{DC}\langle \text{B}\rangle}}
\def\CC{\text{CC}}
\def\CB{\text{CB}}
\def\BB{\text{BB}}
\def\TT{\text{TT}}

\def\nnuc{N_\text{nuc}} %no. of nuclei
\def\nb{N_\text{b}} %no. of spatial basis functions

\def\unittwo{1^{[2]}} %{1_2}
\def\unitfour{1^{[4]}} %{1_4}
\def\four{^{[4]}} %{^{[2]}}
\def\two{{(2)}}
\def\three{{(3)}}

\def\bp{\bos{p}}
\def\br{\bos{r}}

\def\bsigma{\bos{\sigma}}

\def\epsi{\varepsilon}

\def\mL{\Lambda}

\def\pp{{++}}
\def\edc{E^{\pp}_\text{DC}}
\def\edcb{E^{\pp}_\text{DCB}}
\def\pbdc{\langle B \rangle_\DC}
\def\edcpb{E^{\pp}_\text{DC$\langle \text{B}\rangle$}}

\def\nr{\text{nr}}
\def\nonrel{\text{nr}}
\def\reg{\text{rg}}

\usepackage[unicode]{hyperref}
\usepackage{soul}

\definecolor{ao}{rgb}{0.0, 0.5, 0.0}

\newcolumntype{d}[1]{D{.}{.}{#1}}

\usepackage[unicode]{hyperref}
\hypersetup{
   unicode=true,          % non-Latin characters in Acrobat??s bookmarks
   plainpages=false,
   colorlinks=true,       % false: boxed links; true: colored links
   linkcolor=blue,          % color of internal links
   citecolor=blue,        % color of links to bibliography
}

\usepackage{url}

\begin{document}

\title{%
Variational versus perturbative relativistic energies for small and light atomic and molecular systems
}

\author{D\'avid Ferenc} 
\author{P\'eter Jeszenszki} 
\author{Edit M\'atyus} 
\email{edit.matyus@ttk.elte.hu}
\affiliation{ELTE, Eötvös Loránd University, Institute of Chemistry, 
Pázmány Péter sétány 1/A, Budapest, H-1117, Hungary}

\date{\today}

\begin{abstract}
\noindent %
Variational and perturbative relativistic energies are computed and compared for two-electron atoms and molecules with low nuclear charge numbers.
In general, good agreement of the two approaches is observed. 
Remaining deviations can be attributed to higher-order relativistic, also called non-radiative quantum electrodynamics (QED), corrections of the perturbative approach that are automatically included in the variational solution of the no-pair Dirac--Coulomb--Breit (DCB) equation to all orders of the $\alpha$ fine-structure constant. 
The analysis of the polynomial $\alpha$ dependence of the DCB energy makes it possible to determine the leading-order relativistic correction to the non-relativistic energy to high precision without regularization. 
Contributions from the Breit--Pauli Hamiltonian, for which expectation values converge slowly due the singular terms, are implicitly included in the variational procedure.
The $\alpha$ dependence of the no-pair DCB energy shows that  
the higher-order ($\alpha^4\Eh$) non-radiative QED correction is 5~\% of the leading-order ($\alpha^3\Eh$) non-radiative QED correction for $Z=2$ (He), but it is 40~\% already for $Z=4$ (Be$^{2+}$), which indicates that resummation provided by the variational procedure is important already for intermediate nuclear charge numbers. 
\end{abstract}

\maketitle

\section{Introduction \label{sec:intro}}
The non-relativistic quantum electrodynamics framework, which systematically includes all relativistic and quantum electrodynamics (QED) corrections to the non-relativistic energy with increasing powers of the $\alpha$ fine structure constant is the current state of the art for small and light atomic and molecular systems 
\cite{BeSabook57,KinoshitaNio96,Pa06,PazNRQED,PaYePa19,HaZhKoKa2020,AlGiCoKoSc20,KoKaHaZh20,PaYeVlPa21,drake22}. 
Higher precision or higher charge numbers assume the derivation and evaluation of high-order perturbative corrections.

The current state of the art for two-electron systems in the non-relativistic QED framework corresponds to $\alpha^4\Eh$ (in atomic units), which is equivalent to $m\alpha^6$ in natural units \cite{Pa06,PuKoPa17}. 
The $\alpha^5 \Eh$-order corrections have been evaluated for triplet states
of the helium atom \cite{PaYeVlPa21} aiming to resolve current discrepancy of theory and experiment \cite{PaPaYe17,WiPaPuPaYe19,YePaPuPa20,PaYePa21,ClJaScAgScMe21}. The corresponding terms for singlet states have not been completed, yet.

Although the non-relativistic plus relativistic and QED separation has been traditionally (for good reasons) pursued to produce state-of-the-art theoretical values \cite{BuAd17,FeMa19EF,FeMa19HH,PuKoCzPa2019,FeKoMa20,KoKaHaZh20,WeSpPuPa21,PaYePa21,BaKoYaShZh22}, it is possible to partition the relativistic QED problem differently.
The relativistic QED problem of atoms and molecules has two (three) `small' parameters, the $\alpha$ fine structure constant, the $Z\alpha$ nuclear charge number multiple of $\alpha$ (and the electron-nucleus mass ratio, which is not considered in the present work, since the nuclei are fixed). Although $\alpha=1/137.035\ 999\ 084\approx 0.007\ 3$ \cite{codata18} is indeed small, resummation of the perturbation series for $Z\alpha$ would be ideal to cover larger nuclear charge values.

As a starting point for two-electron systems, we consider the Bethe--Salpeter \cite{SaBe51} equation, a relativistic QED wave equation. Following Salpeter's calculation for positronium \cite{Sa52} and Sucher's calculation for the electronic problem of helium \cite{sucherPhD1958}, this equation can be rearranged to an exact equal-times form
\begin{align}
  (H + H_\Delta) \Psi(\bos{r}_1,\bos{r}_2)
  =
  E \Psi(\bos{r}_1,\bos{r}_2),
\end{align}
where $\br_1,\br_2\in\mathbb{R}^3$ are the Cartesian coordinates of the two particles, $H$ is the positive-energy projected two-electron Hamiltonian with instantaneous (Coulomb or Breit) interaction ($I$),
\begin{align}
  H=h_1+h_2 + \mL_{++} I \mL_{++} \; ,
\end{align}
$h_i=c\bos{\alpha}_i\bos{p}_i+\beta m_i c^2 + U 1^{[4]}$ $(i=1,2)$ labels the one-particle Dirac Hamiltonians in the $U$ external Coulomb field of fixed nuclei, and 
\begin{align}
  H_\Delta 
  = 
  \mL_{++} I (1-\mL_{++})
  -
  \mL_{--} I 
  + H_\epsi 
  \label{eq:hdelta}
\end{align}
is a correction term with 
$H_\epsi$, which contains an integral for the relative energy \cite{sucherPhD1958} of two particles, and it carries pair corrections and retardation corrections. Radiative corrections can also be incorporated in $H_\epsi$.
During the derivation \cite{Sa52,sucherPhD1958}, starting out from the interaction of elementary spin-1/2 particles, 
the two-particle (electron) positive-energy Dirac--Coulomb(--Breit) Hamiltonian emerges and $\Lambda_{++}$ ($\Lambda_{--}$) projects onto the positive-(negative-)energy states of two electrons moving in the external field without electron-electron interactions.

Following Sucher, $H_\Delta$ may be considered as perturbation to the positive-energy projected (also called no-pair) Hamiltonian, $H$.
So, the present work is concerned with the numerical solution of the
\begin{align}
  H \Psi = E\Psi
\end{align}
sixteen-component wave equation for the instantaneous Coulomb (C) and Coulomb--Breit (CB) interactions. 
The Breit interaction is either included in the variational solution to obtain the no-pair (++) Dirac--Coulomb--Breit (DCB) energy, $\edcb$, or it is computed as a first-order perturbation to the no-pair Dirac--Coulomb (DC) energy ($\edc$),
\begin{align}
  \langle B \rangle_\DC 
  =
  \langle \Psi_\DC^\pp | B | \Psi_\DC^\pp \rangle \; ,
\end{align}
and it is labelled as $\edcpb = \edc + \pbdc$.

In 1958, Sucher introduced the non-relativistic (Pauli) approximation to the no-pair DC wave function to arrive at numerical predictions.
Using modern computers, we solve the DC and DCB equations numerically to a precision, where comparison with the perturbative treatment (up to the known orders) is interesting and, so far, unexplored. 
This paper is the concluding part of a series of papers \cite{JeFeMa21,JeFeMa22,FeJeMa22}
which report the development of fundamental algorithmic and implementation details of this programme together with the first numerical tests aiming at a parts-per-billion (1:$10^9$) relative precision for the convergence of the variational energy, as well as  comparison with benchmark perturbative relativistic corrections.

\section{Sixteen-component variational methodology \label{sec:method}}
The explicit matrix form of the no-pair Dirac--Coulomb--Breit Hamiltonian for two particles is
\begin{align}
  &H(1,2) = \nonumber \\
  &{ \footnotesize
  \mL_{++}
  \left(%
    \begin{array}{@{} c@{}c@{}c@{}c @{}}
       V\unitfour+U \unitfour & 
       c \bsigma\four_2 \cdot \bp_2 & 
       c \bsigma\four_1 \cdot \bp_1 & 
       B\four \\
       c\bsigma\four_2 \cdot \bp_2 & 
       V\unitfour+(U - 2m_2c^2)\unitfour & 
       B\four & 
       c \bsigma\four_1 \cdot \bp_1 \\
       c\bsigma\four_1 \cdot\bp_1 & 
       B\four &
       V\unitfour+(U-2m_1c^2)\unitfour & 
       c \bsigma\four_2 \cdot \bp_2 \\
       B\four & 
       c \bsigma\four_1 \cdot \bp_1 &
       c \bsigma\four_2 \cdot \bp_2 & 
       V\unitfour+(U-2m_{12}c^2)\unitfour \\
    \end{array}
  \right)
  \mL_{++}
  }
  \label{eq:fullHam}
\end{align}
with $m_{12}=m_1+m_2$, $\bp_i = -\iim(\pd{}{r_{ix}},\pd{}{r_{iy}},\pd{}{r_{iz}})$ ($i=1,2$),
$\bsigma\four_1=(\sigma_x\otimes\unittwo,\sigma_y\otimes\unittwo,\sigma_z\otimes\unittwo)$
and
$\bsigma\four_2=(\unittwo\otimes \sigma_x,\unittwo\otimes\sigma_y,\unittwo\otimes\sigma_z)$, where $\sigma_x,\sigma_y,$ and $\sigma_z$ are the $2\times 2$ Pauli matrices, and $U=\sum_{i=1}^2 \sum_{a=1}^{\nnuc}q_i Z_a/|\bos{r}_i-\bos{R}_a|$ is the external Coulomb potential of the nuclei. 
We note that the operator in Eq.~(\ref{eq:fullHam}) already contains a $-2m_ic^2$ shift ($i=1,2$) for computational convenience and for a straightforward matching of the non-relativistic energy scale.

Regarding the particle-particle interactions in Eq.~(\ref{eq:fullHam}), the Coulomb potential,  
\begin{align}
  V=\frac{q_1q_2}{r_{12}} \; ,
\end{align}
is along the diagonal, 
whereas, the $B\four$ blocks, corresponding to the Breit potential, can be found on the anti diagonal of the Hamiltonian matrix,
\begin{align}
    B\four
    =
    G\four
    -
    \frac{q_1q_2}{2}
    \sum_{i=1}^3\sum_{j=1}^3
      \sigma^{[4]}_{1_i} \sigma^{[4]}_{2_j} 
        \left\lbrace \grad_{1_i}\grad_{2_j}r_{12} \right\rbrace \ .
    \label{eq:breit}
\end{align}
The first term of $B\four$ is called the Gaunt interaction, which reads as 
\begin{align}
     G\four 
     =-\frac{q_1q_2}{r_{12}}\bsigma^{[4]}_1\cdot\bsigma^{[4]}_2
     =-\frac{q_1q_2}{r_{12}}\begin{pmatrix*}[c]
     1 & 0 & 0 & 0 \\
     0 &-1 & 2 & 0 \\
     0 & 2 &-1 & 0 \\
     0 & 0 & 0 & 1
    \end{pmatrix*}   \; .
    \label{eq:gaunt}
\end{align}

The $\mL_{++}$ projector is constructed using the electronic states  (also called positive-energy states, which is to be understood without the $-2m_ic^2$ shift) of Eq.~(\ref{eq:fullHam}) by discarding the $V$ and $B$ particle-particle interaction terms.

We solve the $H\Psi=E\Psi$ wave equation with the no-pair Dirac--Coulomb (Eq.~(\ref{eq:fullHam}) without the $B$ block) and Dirac--Coulomb--Breit operators using a variational-like procedure, a two-particle restricted kinetic balance condition \emph{(vide infra),} and explicitly correlated Gaussian basis functions \cite{JeFeMa21,JeFeMa22,FeJeMa22}.

For a single particle, the (four-component) wave function can be partitioned to large (l, first two) and small (s, last two) components. A good basis representation is provided by the (restricted) kinetic balance condition \cite{Ku84,Liu10}
\begin{align}
    \psi^\text{s}=\frac{\bsigma^{[2]} \bp }{2mc} \psi^\text{l} \; ,
    \label{eq:kb}
\end{align}
connecting the basis function of the small and large components. 
A block-wise direct product  \cite{TrSi72,LiShLi12,ShLiLi17,SiMaRe15,JeFeMa21} is commonly used for the two(many)-particle problem, which is used also in Eq.~(\ref{eq:fullHam}). The corresponding block structure of the two-particle wave function, highlighting the large (l) and small (s) component blocks, is
\begin{align}
    \bos{\Psi}(\br_1,\br_2)=\left( \begin{array}{c}
         \psi^\text{ll}(\br_1,\br_2)  \\
         \psi^\text{ls}(\br_1,\br_2)  \\
         \psi^\text{sl}(\br_1,\br_2)  \\
         \psi^\text{ss}(\br_1,\br_2)  
    \end{array} \right)    \; .
\end{align}
We have implemented \cite{JeFeMa21,JeFeMa22,FeJeMa22} the simplest two-particle generalization of the restricted kinetic balance, Eq.~(\ref{eq:kb}), in the two-particle basis set in the sense of a transformation or metric \cite{Ku84}:
\begin{align}
    H_\text{KB}=X^\dagger H X \; , 
    && 
    X=\text{diag}\left(%
      1\four, \frac{\left(\bsigma_2\four \bp_2 \right)}{2m_2c},%
      \frac{\left(\bsigma_1\four \bp_1 \right)}{2m_1c},%
      \frac{\left(\bsigma_1\four \bp_1 \right)\left(\bsigma_2\four \bp_2 \right)}{4m_1m_2c^2}\right) \; .
    \label{eq:XHX} 
\end{align}
The transformed operators for the DC and the DCB problem are given in Refs.~\cite{JeFeMa22} and \cite{FeJeMa22}, respectively.

The sixteen-component wave function is written as a linear-combination of anti-symmetrized \cite{JeFeMa21} spinor functions 
\begin{align}
  \Psi(\br_1,\br_2) 
  &=
  \mathcal{A} \sum_{i=1}^{\nb} \sum_{\chi=1}^{16} c_{i\chi} \bos{d}_{\chi}
   \Theta_i\left(\br_1,\br_2;\bos{A}_i,\bos{s}_i\right) \; ,
\end{align}
where the spatial part is represented by explicitly correlated Gaussians functions (ECGs),
\begin{align}
  \Theta_i\left(\br_1,\br_2;\bos{A}_i,\bos{s}_i\right) 
  &= 
  \exp\left[ -(\br-\bos{s}_i)^T\left(\bos{A}_i\otimes 1^{[3]}\right)(\br-\bos{s}_i)\right]  \; .
     \label{eq:ECGbasis}
\end{align}

For low $Z$ systems, it is convenient to work in the $LS$ coupling scheme. We optimized the ECG parameterization for the ground (and first excited state) by minimization of the ground (and first excited) totally symmetric, non-relativistic singlet energy. 
To be able to generate (relatively) large basis sets and to achieve good (parts-per-billion relative) convergence of not only the non-relativistic, but also the $\DCpB$ and DCB energies,
the value of the energy functional, which we minimized, was incremented by a `penalty' term \cite{TuPaAd10,BuAd20} that helped us to generate and optimize ECG basis functions that are less linearly dependent (and thus, well represented in double precision arithmetic). 
The same basis set was used to construct the non-interacting problem, Eq.~(\ref{eq:fullHam}) without $V$ and $B$, and the positive-energy projector. We used the cutting projection approach and checked some of the results with the complex scaling (CCR) projector \cite{JeFeMa22}.
The triplet contributions are estimated to be small (in perturbative relativistic computations, they appear at $\alpha^4\Eh$ order \cite{Pa06,PuKoPa17}), and will be reported for the present framework in the future.

All computations have been carried out with an implementation of the outlined algorithm (see also  Refs.~\cite{JeFeMa21,JeFeMa22,FeJeMa22}) in the QUANTEN computer program, used in pre-Born--Oppenheimer, non-adiabatic, and (regularized) perturbative relativistic computations \cite{Ma19review,FeMa19EF,FeMa19HH,MaFe22,FeKoMa20,JeIrFeMa22}.
Throughout this work, Hartree atomic units are used and the speed of light is $c=\alpha^{-1}a_0\Eh/\hbar$ with $\alpha^{-1}=137.$035 999 084  \cite{codata18}.

\section{Comparison of the perturbative and variational energies}
The Dirac--Coulomb--Breit energies, $\edcpb$ and  $\edcb$, obtained from sixteen-component computations in this work are compared with perturbation theory results precisely evaluated with well-converged non-relativistic wave functions (taken from benchmark literature values).

The aim of this comparison is three-fold.

(a) First, it is a numerical check, whether the sixteen-component approach can reproduce the established perturbative benchmarks with a parts-per-billion (ppb) precision, which 
corresponds to an energy resolution that is relevant for the current experiment-theory comparison.

(b) Second, it is about understanding the variational results. The sixteen-component variational computation includes a `resummation' of the perturbation series in $Z\alpha$ for part of the problem. 
Identification of the relevant higher-order perturbative corrections provides an additional check for the implementation and for a good understanding of the developed variational relativistic methodology.

(c) Third, after completion of (a) and (b), we may estimate the importance of missing orders of the perturbative approach, since the sixteen-component computation provides all $Z\alpha$ orders for the relevant part of the problem.

The present comparison provides a starting point for further developments aiming at the inclusion of missing `effects', in particular, contributions from the $H_\Delta$ term in Eq.~(\ref{eq:hdelta}), including pair correction, retardation, etc.,  as well as radiative corrections and motion of the nuclei.

\subsection{Perturbative energy expressions}
The leading, $\alpha^2\Eh$, order, often called `relativistic correction' to the non-relativistic energy is obtained by a perturbative approach, either by the Foldy--Wouthuysen transformation \cite{DyFaBook07} or by Sucher's approach \cite{sucherPhD1958} (in some steps reminiscent of the later Douglas--Kroll transformation \cite{DoKr74}) for the Dirac--Coulomb (DC) and Dirac--Coulomb--Breit (DCB) Hamiltonians.
The energy up to second order in $\alpha$ (in atomic units) reads as
\begin{align}
  E_\DCB^{(2)}
  =
  E_\text{nr}^{(0)}
  +
  \alpha^2
  \langle
    \Psi_\text{nr} |
      H_\DCB^{(2)}
    |\Psi_\text{nr}
  \rangle
  =
  E_\text{nr}^{(0)}
  +
  \alpha^2
  \langle
    H_\DC^{(2)}+H_\text{B}^{(2)}
  \rangle_\nonrel \; ,
  \label{eq:ptalpha2}
\end{align}
where $\Psi_\text{nr}$ is the non-relativistic wave function and the 
$\langle O\rangle_\nonrel=\langle \Psi_\nonrel | O | \Psi_\nonrel \rangle$
short notation is introduced. 
Furthermore, 
\begin{align}
  H_\DC^{(2)}
  &=    
  -\frac{1}{8}\sum_{i=1}^2 (\bos{\nabla}_i^2)^2
  + 
  \frac{\pi}{2}\sum_{i=1}^2\sum_{A=1}^{N_\mathrm{nuc}} Z_A \delta(\br_{iA}) 
  -\pi \sum_{i=1}^2\sum_{j>i}^2 \delta(\br_{ij}) \; ,
  \label{eq:fwptdc}
  \\
  H_\text{B}^{(2)}
  &=
  H_\text{OO} + 2\pi \sum_{i=1}^2\sum_{j>i}^2 \delta(\br_{ij})
  \label{eq:fwptbreit}
\end{align}
with
\begin{align}
  H_\text{OO}
  =
  -\sum_{i=1}^2 \sum_{j>i}^2 
  \frac{1}{2r_{ij}}
  \left(%
    \bos{p}_i\bos{p}_j
    +
    \frac{\bos{r}_{ij}(\bos{r}_{ij}\bos{p}_i)\bos{p}_j}{r_{ij}^2}
  \right)  \; .
  \label{eq:fwptoo}
\end{align}

To obtain precise correction values, regularization techniques \cite{Dr81,PaCeKo05,JeIrFeMa22} have been used to pinpoint the value of the non-relativistic expectation value of the singular terms, $\delta(\br_{iA})$, $\delta(\br_{ij})$, and $(\bos{\nabla}^2_i)^2$.

Furthermore, we have noticed in earlier work \cite{JeFeMa22,FeJeMa22} that the `non-radiative QED' corrections of the perturbative scheme are `visible' at the current ppb convergence level already for $Z=1$. For this reason, we collect the relevant positive-energy corrections from Sucher's work \cite{sucherPhD1958} in the following paragraphs. 
It is important to point out that these terms contribute to the $\alpha^3\Eh$-order perturbative corrections, but provide only part of the full correction at this order, which was first derived by Araki (1957) \cite{araki57} and Sucher  (1958) \cite{sucherPhD1958}.

\vspace{0.25cm}
\paragraph{Leading-order non-radiative QED corrections ($\alpha^3\Eh$) to the no-pair energy} ~\\
The two-Coulomb-photon exchange correction (p.~52, Eq.~(3.99) \cite{sucherPhD1958}) is
\begin{align}
  \varepsilon_\text{CC}^{++}
  = 
  -\left(\frac{\pi}{2}+\frac{5}{3}\right) \langle \delta(\bos{r}_{12})\rangle_\nr
  \approx
  -3.237\ \langle \delta(\bos{r}_{12})\rangle_\nr \; .
  \label{eq:ccpp}  
\end{align}
The correction due to one (instantaneous) Breit photon exchange with resummation for the Coulomb ladder (p.~80, Eq.~(5.64) \cite{sucherPhD1958}) is
\begin{align}
  \varepsilon_\text{CB}^{++}
  =
  4\left(\frac{\pi}{2}+1\right) \langle \delta(\bos{r}_{12})\rangle_\nr
  \approx 
  10.283\ \langle \delta(\bos{r}_{12})\rangle_\nr \; .
  \label{eq:cbpp}
\end{align}
We note that this correction corresponds to the unretarded (Breit) part of the transverse photon exchange (Ref.~\cite{sucherPhD1958} uses the Coulomb gauge for this part), and the retardation correction to this term is evaluated separately (not considered in this work).

Finally, the correction due to the exchange of two (retarded) transverse photons according to Sucher (p.~93, Eq.~(6.9b++)  \cite{sucherPhD1958}) is
\begin{align}
  \varepsilon_\text{TT}^{++}
  =
  -\frac{\pi}{2} \langle \delta(\bos{r}_{12})\rangle_\nr
  \approx
  -1.571\ \langle \delta(\bos{r}_{12})\rangle_\nr \; .
  \label{eq:ttpp}
\end{align}
It is necessary to note that this term includes retardation effects, whereas our computation, does not. 
For this reason, the comparison of this term with the results of the variational $\DCB$ computation is only approximate and not fully quantitative.

In summary, the following $\alpha^3\Eh$-order perturbative energies will be used for comparison
with the $\DCpB$\ and $\DCB$\ sixteen-component computations,
\begin{align}
  E^{(3)}_{\text{DC}\langle\text{B}\rangle}
  =
  E^{(2)}_\text{DCB}
  +
  \alpha^3 (\varepsilon_\text{CC}^{++}+\varepsilon_\text{CB}^{++})
  &=
  E^{(2)}_\text{DCB} + \alpha^3 \left(\frac{3 \pi}{2} + \frac{7}{3}\right)
  \langle \delta(\bos{r}_{12})\rangle_\nr 
  \label{eq:ptDCpB} \\
  &\approx
  E^{(2)}_\text{DCB}+ \alpha^3\ 7.046\ \langle \delta(\bos{r}_{12})\rangle_\nr \; ,
  \nonumber 
\end{align}
where we note that $E_\text{DCB}^{(2)}=E_{\text{DC}\langle\text{B}\rangle}^{(2)}$ ($\alpha^2\Eh$ order), and
\begin{align}
  E^{(3)}_\text{DCB}
  \approx
  E^{(2)}_\text{DCB}
  +
  \alpha^3 (\varepsilon_\text{CC}^{++}+\varepsilon_\text{CB}^{++}+\varepsilon_\text{TT}^{++})
  &=
  E^{(2)}_\text{DCB}
  +
  \alpha^3\left(\pi + \frac{7}{3}\right) \langle\delta(\bos{r}_{12})\rangle_\nr
  \label{eq:ptDCB}  
  \\
  &\approx 
  E^{(2)}_\text{DCB}+ \alpha^3\ 5.475\ \langle \delta(\bos{r}_{12})\rangle_\nr \; ,
  \nonumber 
\end{align}
where, in particular, the single Breit and two-transverse corrections sum to
\begin{align}
  (\epsi_{\CB}^\pp + \epsi_{\TT}^\pp)\ \alpha^3 
  =
  \left(%
    4 + \frac{3\pi}{2}
  \right)
  \langle \delta(\bos{r}_{12})\rangle_\nr\   \alpha^3 
  \approx
  8.712\ 
  \langle \delta(\bos{r}_{12})\rangle_\nr\  \alpha^3\  \; .
  \label{eq:cbtt}
\end{align}

\begin{table}
\caption{%
    The no-pair DC energy with first-order perturbative Breit correction, 
    $E_{\DC\langle \text{B}\rangle}^\pp$ in $\Eh$, and
    the no-pair DCB energy, $E_\DCB^\pp$ in $\Eh$.
    The differences from the $\alpha^2\Eh$- and $\alpha^3\Eh$-order perturbative energies, $\delta^\two$ and $\delta^\three$ in n$\Eh$, are also shown. 
    Ground-state energies are reported unless otherwise indicated.
    For up to $Z=2$ systems, all digits are significant, for Li$^+$ (Be$^{2+}$)
    the last one (two) digits are estimated to be uncertain.
    \label{tab:compare}
}
  \begin{tabular}{@{}l r@{}l@{} d{3.9}@{\ \ }d{5.1}@{\ \ }d{3.1}@{\ \ \ \ \ } d{3.9}@{\ \ }d{5.1}@{\ \ }d{4.1}@{\ \ }d{3.1}@{}}
    \hline\hline\\[-0.35cm]
    \multicolumn{1}{c}{} &
     & &
    \multicolumn{1}{c}{$E_{\DCpB}^\pp$} &
    \multicolumn{1}{c}{${\delta^\two_{\DCpB}} ^\mathrm{a}$} &
    \multicolumn{1}{c}{${\delta^\three_{\DCpB}} ^\mathrm{a}$} &
    \multicolumn{1}{c}{$E_\DCB^\pp$} &
    \multicolumn{1}{c}{${\delta^\two_{\DCpB}} ^\mathrm{a}$} & 
    \multicolumn{1}{c}{${\delta^\three_{\DCpB}} ^\mathrm{a}$} & 
    \multicolumn{1}{c}{${\delta^\three_{\DCB}} ^\mathrm{a}$} \\    
    \hline\\[-0.35cm]
    H${_2} ^\mathrm{b}$ 
    &&
    & -1.174\ 486\ 622 & 45 & 1
    & -1.174\ 486\ 635 & 32 & -14 & -4  \\
    H${_3^+} ^\mathrm{b}$ 
    &&  
    & -1.343\ 847\ 366  & 50 & 0
    & -1.343\ 847\ 381  & 35 & -21 & -4 \\
    HeH${^+} ^\mathrm{b}$ 
    & &  
    & -2.978\ 807\ 919 & 261 & -16
    & -2.978\ 808\ 003  & 177 & -100 & -38  \\    
    \hline\\[-0.35cm]
    H$^-$ &&                
    & -0.527\ 756\ 279 & 74 & 0
    & -0.527\ 756\ 281 & 5 & -2 & -1 \\
    He ($2S$) $^\mathrm{c}$
    && 
    & -2.146\ 082\ 355   & 13 & -11
    & -2.146\ 082\ 363   & 5 & -19 & -13 \\
    He ($1S$) $^\mathrm{c}$ 
    &&                
    & -2.903\ 828\ 032 & 279 & -13
    & -2.903\ 828\ 121 & 190 & -102 & -37 \\
    Li$^+$       &&                
    & -7.280\ 540\ 978 & 1300 & -161
    & -7.280\ 541\ 443 & 835 & -626 & -301 \\
    Be$^{2+}$       &&                
    & -13.657\ 788\ 729  & 3175 & -995
    & -13.657\ 790\ 100  & 1804 & -2366 & -1436 \\
    \hline\hline
  \end{tabular}
  \begin{flushleft}
    $^\mathrm{a}$~$\delta^{(n)}_x=E^\pp_x-E^{(n)}_x$ with $n=2,3$ and $x$ stands for $\DCpB$ or $\DCB$. The expressions for $E^{(n)}_x$  are listed in Eqs.~(\ref{eq:ptalpha2})--(\ref{eq:ptDCB})
    and the reference non-relativistic energy and integral values are collected in Table~S13.
    We note that $\delta^\two_{\DCpB}=\delta^\two_{\DCB}$. \\
    $^\mathrm{b}$~Electronic ground state for the nuclear-nuclear distance,
    $R_\text{eq}=1.4$~bohr, 1.65~bohr, and 1.46~bohr for H$_2$, H$_3^+$, and HeH$^+$, respectively. \\
    $^\mathrm{c}$~$1S$ and $2S$ are used for 1 and $2\ ^1S_0$, respectively. 
  \end{flushleft}
\end{table}

\subsection{Sixteen-component, variational results}
Table \ref{tab:compare} shows the sixteen-component, no-pair $\DCpB$ and $\DCB$ energies computed in this work 
and their comparison 
with the $\alpha^2\Eh$- and $\alpha^3\Eh$-order perturbative results. 
The $\DCpB$ and $\DCB$ energies reported in this table differ from our earlier work \cite{JeFeMa21,FeJeMa22}. (The entries of the earlier reported tables for the Breit energies were in an error due to a programming mistake during the construction of the sixteen-component submatrices for pairs of ECG functions. It did not affect the DC (singlet) energies \cite{JeFeMa22}, but affected the energies including the Breit interaction \cite{FeJeMa22}.)

According to Table~\ref{tab:compare}, the deviation of the variational results from the $\alpha^2\Eh$-order energies is on the order of a few 10~n$\Eh$ for $Z=1$, whereas it is a few 100~n$\Eh$ already for $Z=2$. 
For a better comparison, it is necessary to include the $\alpha^3\Eh$-order (non-radiative QED) corrections to the perturbative energy.
The relevant terms correspond to  the two-Coulomb-photon, $\epsi_\CC^\pp$, Eq.~(\ref{eq:ccpp}), the Coulomb-Breit-photon, $\epsi_\CB^\pp$, Eq.~(\ref{eq:cbpp}), 
and the Breit-Breit-photon exchange corrections (for the positive-energy states). The last correction can be approximated with the (more complete) two-transverse photon exchange correction, Eq.~(\ref{eq:ttpp}) (that is available from \cite{sucherPhD1958}). 

It was shown in Ref.~\cite{JeFeMa22} that inclusion of the $\alpha^3\Eh$-order, positive-energy Coulomb-ladder correction, $\epsi_\CC^\pp$, in the perturbative energy closes the gap between the perturbative and variational energies for the lowest $Z$ values.
Most importantly for the present work, inclusion of $\epsi_\CC^\pp$ and $\epsi_\CB^\pp$, reduces the deviation for the no-pair $\DCpB$ energy from the perturbative value to near 0 n$\Eh$ for $Z=1$ and to ca. 10~n$\Eh$\ for $Z=2$. 

Regarding the variational $\DCB$ energy, there is a non-negligible remainder between the variational, $\edcb$, and perturbative energies, $E_{\DCB}^{(3)}$, Eq.~(\ref{eq:ptDCB}), a few n$\Eh$ for $Z=1$ and a few tens of n$\Eh$ for $Z=2$.
This small, remaining deviation must be due to the fact that the sixteen-component computation reported in this work 
(a) does not include retardation, but
(b) includes `effects' beyond $\alpha^3\Eh$ order.

First of all, $E_\DCB^\pp$ includes the exchange of two and more (unretarded) Breit photons. 
To constrain the number of Breit photon exchanges, instead of the variational DCB computation, we can consider perturbative corrections due to the Breit interaction
to the sixteen-component DC wave function. 
The first-order correction,
\begin{align}
    \langle B \rangle_{\DC,n}
    &=
    \langle %
      \Psi^{\pp}_{\text{DC},n} |  X^\dagger B(1,2) X | \Psi^\pp_{\text{DC},n}
    \rangle \; , 
    \label{eq:pt1}
\end{align}
corresponds to a single Breit-photon exchange (in addition to the Coulomb ladder), 
while the first- and second-order perturbative corrections  \cite{FeJeMa22}, 
\begin{align}    
    \mathcal{P}^{(2)}_n\lbrace B\rbrace
    &=
    \langle B \rangle_{\DC,n}
    +
    \sum_{i \neq n} 
    \frac{%
      \abs{%
        \langle%
          \Psi^{\pp}_{\text{DC},i} 
          | X^\dagger B(1,2) X | 
          \Psi^{{\pp}}_{\text{DC},n}
        \rangle
      }^2
    }{%
      E^{\pp}_{\text{DC},i}-E^{\pp}_{\text{DC},n}
    }  
    \label{eq:pt2}
    \; ,
\end{align}
account for the effect of one- and two- (non-crossing) Breit photons.

We have numerically observed that $E_{\DC}^\pp + \mathcal{P}^{(2)}_n\lbrace B\rbrace$ reproduces $\edcb$
within a few n$\Eh$ (Tables~S1--S8), which indicates that the energy is dominated by at most two Breit-photon exchanges in all systems studied in this work (up to $Z=4$). 

In Table~\ref{tab:compare}, a relatively good agreement can be observed with the $\alpha^3\Eh$-order perturbative energies for $Z=1$ and $2$, but we observe a larger deviation between the sixteen-component and perturbative results for $Z=3$ and 4, which indicates that inclusion of higher-order perturbative corrections would be necessary for a good (better) agreement. 
The non-radiative, singlet part of the $\alpha^4\Eh$ correction (after cancelling divergences) has been reported for both  He (1S) and (2S) to be $-$11~n$\Eh$\ \cite{Pa06}, which is in an excellent agreement with the $\delta_{\DCpB}^{(3)}=-11$ and $-13$~n$\Eh$\ values in Table~\ref{tab:compare}. It is necessary to note that the comparison is only approximate, since the perturbative value contains contributions also from other `effects'. 
We note that $\alpha^4\Eh$-order computations have been reported in Ref.~\cite{YePa10} 
for $Z=3$ and 4 (Li$^+$ and Be$^{2+}$) ground states,  
but we could not separate the non-radiative QED part from the given data.

To disentangle the contribution of the different $\alpha$ orders, and hence, to have a more direct comparison with the perturbative results, we have studied the $\alpha$ dependence of $\edcpb$ and $\edcb$.

\begin{table}%[h]
  \caption{%
  H$^-$ ($1S$): 
  Convergence of the quartic $\alpha$ polynomial coefficients (only significant digits shown) obtained by fitting to sixteen-component energy computations (see also Figs.~\ref{fig:BreitPTalphafit} and \ref{fig:DCBalphafit} of this work and Fig.~3 of Ref.~\cite{JeFeMa22}).
  $N_\text{b}$ is the number of the ECG basis functions.
  \label{tab:hmFitParamConv}
  }
  \begin{tabular}{@{}l @{\hspace{-0.5cm}}c d{3.5}d{3.3}d{4.1} c d{2.5}d{2.3}d{3.1} c d{2.6}d{2.3}d{3.1} @{}}
    \hline\hline\\[-0.35cm]
    & &
    \multicolumn{3}{c}{%
     \scalebox{1.}{$\frac{E_\DC^\pp-\alpha^2\langle H_\DC^{(2)}\rangle_\nonrel}{\langle\delta(\bos{r}_{12})\rangle_\nonrel}$}
    }  & &
    \multicolumn{3}{c}{%
      \scalebox{1.}{$\frac{\langle B \rangle_\DC - \alpha^2\langle H_\text{B}^{(2)}\rangle_\nonrel}{\langle\delta(\bos{r}_{12})\rangle_\nonrel}$}
    } & &
    \multicolumn{3}{c}{%
     \scalebox{1.}{$\frac{E_\DCB^\pp - E_\DC^\pp - \alpha^2\langle H_\text{B}^{(2)}\rangle_\nonrel}{\langle\delta(\bos{r}_{12})\rangle_\nonrel}$}
    } \\[0.15cm]
    \footnotesize{fit:} & &
    \multicolumn{3}{c}{%
      \footnotesize{$c_2\alpha^2 + c_3\alpha^3 + c_4\alpha^4$}
    }  & &
    \multicolumn{3}{c}{%
      \footnotesize{$b_2\alpha^2 + b_3\alpha^3 + b_4\alpha^4$}
    } & &
    \multicolumn{3}{c}{%
      \footnotesize{$\tilde{b}_2\alpha^2 + \tilde{b}_3\alpha^3 + \tilde{b}_4\alpha^4$}
    }    
    \\[0.03cm]
    \cline{3-5}\cline{7-9}\cline{11-13}\\[-0.33cm]
    \multicolumn{1}{l}{$N_\mathrm{b}$} & &
    \multicolumn{1}{c}{$c_2$}  &
    \multicolumn{1}{c}{$c_3$}  &
    \multicolumn{1}{c}{$c_4$}  & &
    \multicolumn{1}{c}{$b_2$}  &
    \multicolumn{1}{c}{$b_3$}  &
    \multicolumn{1}{c}{$b_4$}  & &
    \multicolumn{1}{c}{$\tilde{b}_2$}  &
    \multicolumn{1}{c}{$\tilde{b}_3$}  &
    \multicolumn{1}{c}{$\tilde{b}_4$}
    \\
    \hline\\[-0.35cm]
    300	&& 0.0034	& -0.9   & -115  && -0.0094   & 3.25     & 264 && -0.0051   & 1.72  & 228 \\
    400	&& 0.00853	& -3.32  & -19.7 && -0.01479  & 9.891    & 7.0 && -0.01359  & 7.534 & 31.9 \\
    500	&& 0.00001  & -3.262 & -23.1 && -0.00815  & 10.050   & 5.7 && -0.008113 & 7.878 & 21.5 \\
    600	&& -0.00003 & -3.26  & -23.2 && -0.00846  & 10.049   & 5.7 && -0.008106 & 7.877 & 21.5 \\
    \hline\\[-0.35cm]
    Sucher \cite{sucherPhD1958} $^\text{a}$
        &&               & -3.237  &         &&             & 10.283 &     &&             & 8.712 &  \\
    $\Delta$ $^\text{b}$  
        &&               & -0.02  &         &&             & -0.234 &     &&             & -0.835 &  \\
    \hline\hline
    \end{tabular}
    \begin{flushleft}
    {\footnotesize %\\
    $^\text{a}$ %
    Ref.~\cite{sucherPhD1958}:
    $c_{3,\text{Su58}}=-\left(\frac{\pi}{2} + \frac{5}{3} \right) \approx -3.237$;
    $b_{3,\text{Su58}}=4\left(\frac{\pi}{2} + 1 \right) \approx 10.283$;
    $\tilde{b}_{3,\text{Su58}}=4+\frac{3\pi}{2} \approx 8.712$. \\
    $^\text{b}$ %
    $\Delta$: deviation of the $\alpha^3\Eh$-order coefficient of this work and of Sucher's analytic expressions~\cite{sucherPhD1958}.}
    \end{flushleft}    
\end{table}

\section{Fine-structure constant dependence of the Dirac--Coulomb--Breit energies}
The sixteen-component $\DCpB$ and $\DCB$ computations have been repeated with slightly different values used for the $\alpha$ coupling constant of the electromagnetic interaction (Figs.~\ref{fig:BreitPTalphafit} and \ref{fig:DCBalphafit}). A quartic polynomial of $\alpha$ was fitted to the computed series of $\edcpb(\alpha)$ and $\edcb(\alpha)$ energies.
The fitted coefficients of the $\alpha$ polynomial can be directly compared with the perturbative corrections, Eqs.~(\ref{eq:ptalpha2})--(\ref{eq:ptDCB}), corresponding to the same order in $\alpha$.
We call this approach the $\alpha$-scaling procedure (to the analysis of the sixteen-component results).

\subsection{Comparison of the $\alpha^3\Eh$ contributions}
For a start, we note that the `$\alpha$-scaling' procedure was already successfully used for $\edc(\alpha)$ in Ref.~\cite{JeFeMa22} and resulted in two important observations. First, the $\alpha^3\Eh$-order term of the polynomial fitted to the $\edc(\alpha)$ points,
\begin{align}
  c_3\ \alpha^3 = -3.26(1)\ \langle\delta(\bos{r}_{12})\rangle_\text{nr}\ \alpha^3 \; ,
\end{align}
is in an excellent agreement with Sucher's positive-energy two-Coulomb-photon correction,
$\varepsilon_\text{CC}^{++}$, Eq.~(\ref{eq:ccpp}) \cite{sucherPhD1958}.

Regarding the Breit term, a similar $\alpha$-scaling procedure resulted in contradictory observations in Ref.~\cite{FeJeMa22}. It has turned out very recently that the contradictory observations were caused by a programming error in the construction of the DC(B) matrix. After noting and correcting this error, we have recomputed all reported values. The (singlet) DC energies \cite{JeFeMa21,JeFeMa22} did not change, but the $\DCpB$ and $\DCB$ energies \cite{JeFeMa21,JeFeMa22} were affected. In the present work, we report the recomputed values and enlarged the basis sets for some of the systems, so it is possible now to achieve a (sub-)ppb-level of relative precision for the variational energy including the Breit interaction.

\begin{figure}
    \centering
    \includegraphics[scale=1]{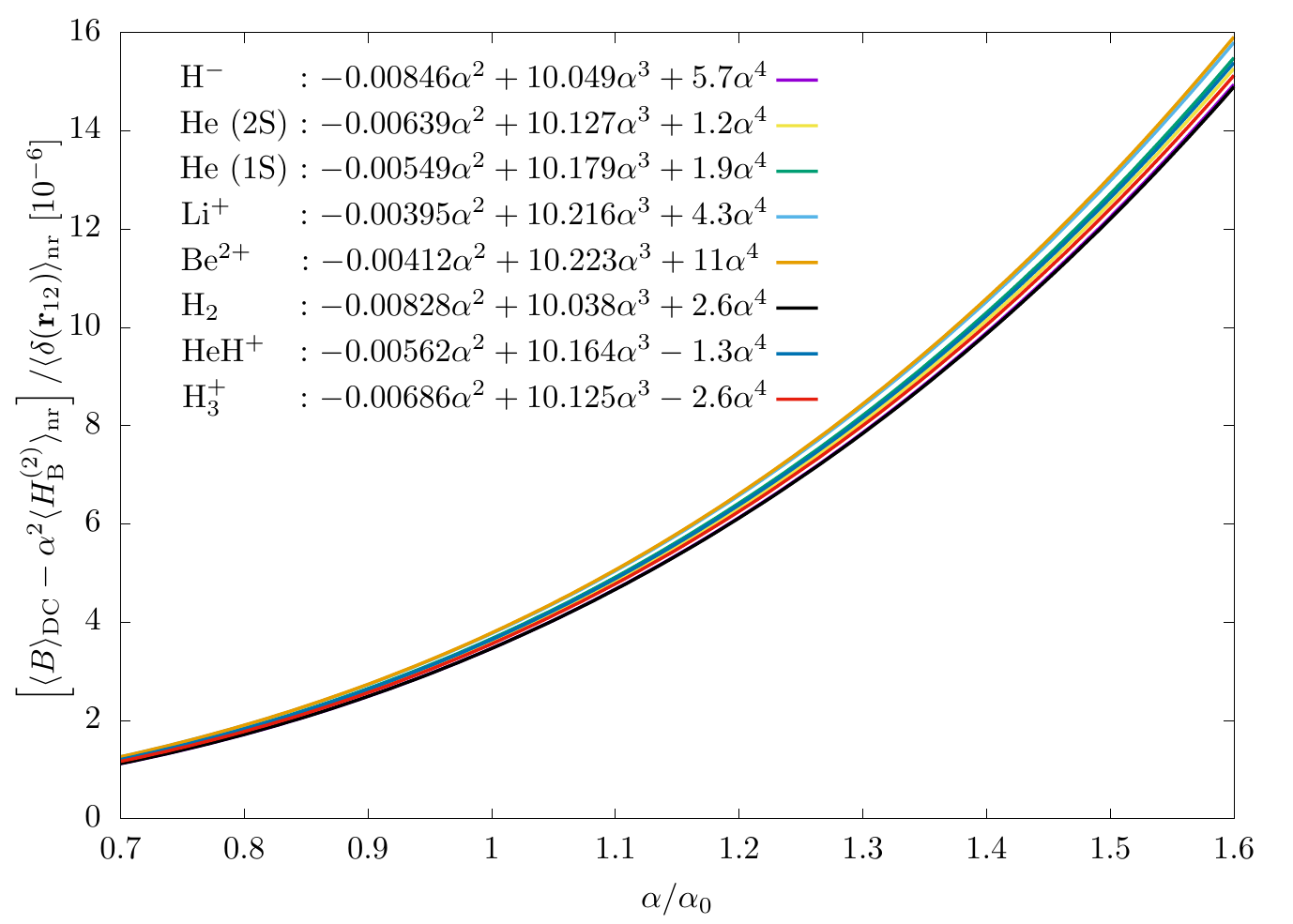}
    \caption{%
      Dependence of the Breit correction to the no-pair DC energy, $\langle B\rangle_\text{DC}$, on the $\alpha$ coupling constant of the electromagnetic interaction. 
      Hartree atomic units are used and 
      $\alpha_0$
      labels $1/137.$035~999~084 \cite{codata18}. 
      The data points, used for fitting the polynomials, were computed at the 
      $\alpha=1/(\alpha_0+n)$, $n=-50, \dots, 50$ values.      
      The $\langle H_\text{B}^{(2)} \rangle_\mathrm{nr}$ and 
      $\langle \delta(\bos{r}_{12})\rangle_\text{nr}$ values used to prepare this figure were calculated by `direct' integration (without regularization, Table~S14).
      \label{fig:BreitPTalphafit}
    }
\end{figure}

First of all, both the $\edcpb(\alpha)$ and $\edcb(\alpha)$ data points (corresponding to a series of slightly different $\alpha$ values) can be fitted well with a quartic polynomial of $\alpha$. 
For practical numerical reasons, we did not fit a general, fourth-order polynomial directly to $\edcpb(\alpha)$, but we subtracted a good approximate value for the `large' leading-order ($\alpha^2\Eh$) relativistic correction, \emph{i.e.,} $\langle B\rangle_\DC - \alpha^2 \langle H^{(2)}_\text{B} \rangle_\text{nr}$. 
We emphasize that the second term, $\alpha^2\langle H_\text{B}^{(2)}\rangle_\nonrel$, is a simple quadratic function of $\alpha$, since the
$\langle H_\text{B}^{(2)}\rangle_\nonrel$
non-relativistic Breit correction is independent of $\alpha$ in atomic units. 
To bring the several systems studied in this work to the same scale, we divided the difference by $\langle \delta(\br_{12})\rangle_\text{nr}$ (Fig.~\ref{fig:BreitPTalphafit}).
For the generation of the figure and the $\alpha$ polynomials, we used  the $\langle H^{(2)}_\text{B} \rangle_\text{nr}$ and $\langle \delta(\br_{12})\rangle_\text{nr}$ values evaluated `directly' (without regularization) in the (largest) spatial ECG basis set used for the sixteen-component computations (Table~S14).

\begin{figure}
    \centering
    \includegraphics[scale=1]{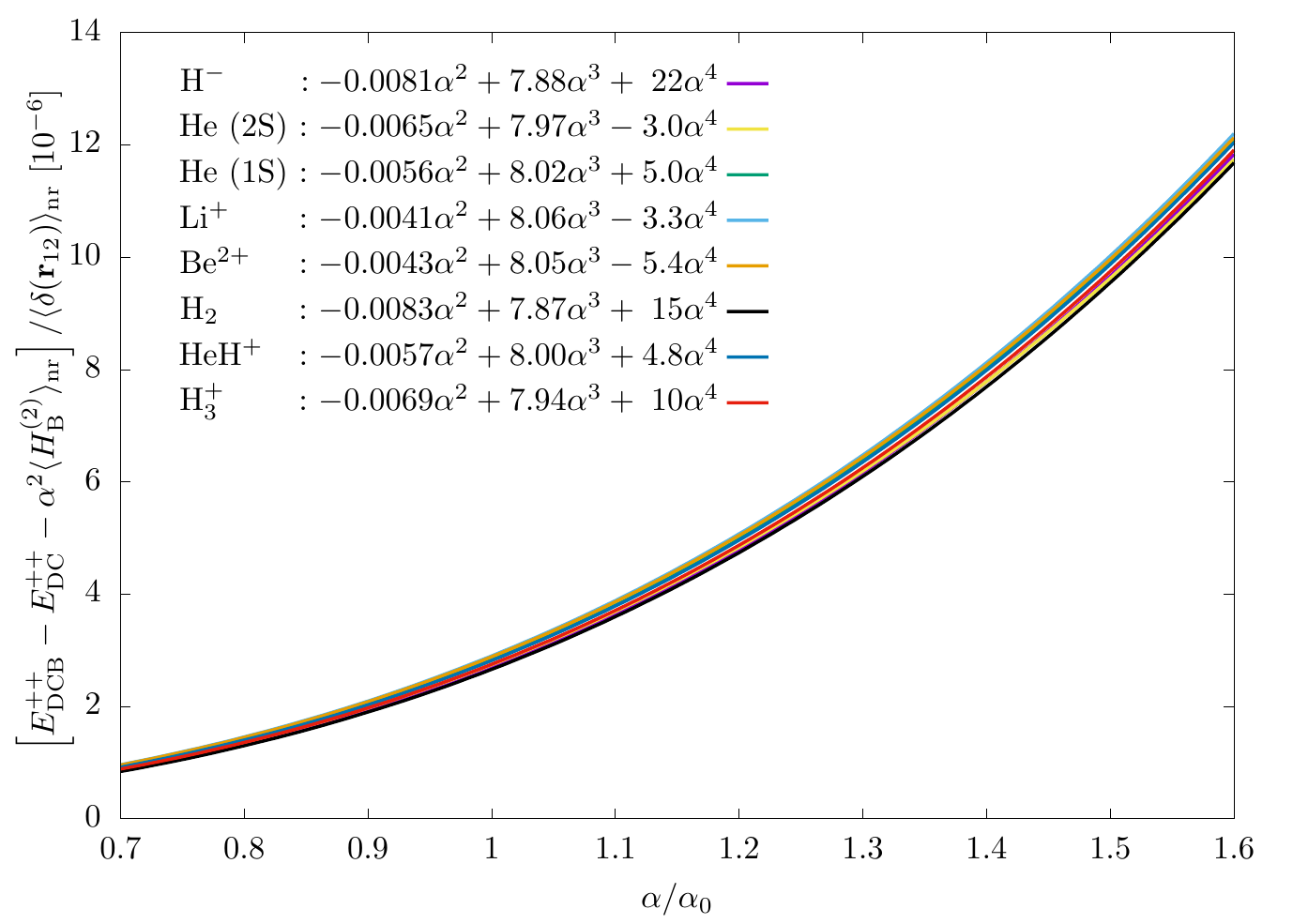}
    \caption{%
      Dependence of the no-pair DCB and DC energy difference on the $\alpha$ coupling constant of the electromagnetic interaction. 
      Hartree atomic units are used and $\alpha_0$ labels $1/137.$035~999~084 \cite{codata18}. 
      The data points, used for fitting the polynomials, were computed at the $\alpha=1/(\alpha_0+n)$, $n=-50, \dots, 50$ values.      
      The $\langle H_\text{B}^{(2)} \rangle_\mathrm{nr}$ and
      $\langle \delta(\bos{r}_{12})\rangle_\text{nr}$ values used to prepare this figure were calculated by `direct' integration (without regularization, Table~S14).
      \label{fig:DCBalphafit}
    }
\end{figure}

From a computational point of view, it was the most difficult to have a stable $\alpha$ fit for H$^-$ in double precision arithmetic (due to the smallness of $\langle \delta(\bos{r}_{12})\rangle_\nonrel$), and for this reason, we show the convergence of the fitted coefficients with respect to the basis set size in Table~\ref{tab:hmFitParamConv}. For all other systems studied in this work, the fitting was numerically more robust.

The $\alpha^3\Eh$-order term in $\langle B\rangle_\DC(\alpha)$ (Fig.~\ref{fig:BreitPTalphafit}) is obtained from the cubic term of the fit, 
\begin{align}
  b_3\ \langle \delta(\br_{12})\rangle_\text{nr}\ \alpha^3 
  = 
  10.1(2)\ \langle \delta(\br_{12})\rangle_\text{nr}\ \alpha^3 \; ,
\end{align}
which is in an excellent agreement with the perturbative correction due to a single Breit photon in addition to the Coulomb ladder (for the positve-energy states), 
$\epsi_\CB^\pp\ \alpha^3$, Eq.~(\ref{eq:cbpp}).
As to $\edcb(\alpha)$ (Fig.~\ref{fig:DCBalphafit}),  we obtain the $\alpha^3\Eh$-order term as
\begin{align}
  \tilde{b}_3\ \langle \delta(\br_{12})\rangle_\text{nr}\ \alpha^3 
  = 
  8.98(25)\ \langle \delta(\br_{12})\rangle_\text{nr}\ \alpha^3 \; .
  \label{eq:tbthree}
\end{align}
This value can be compared with the $\alpha^3\Eh$-order positive-energy correction of
the one- and two-Breit photon exchanges (in addition to the Coulomb ladder), 
$\epsi_{\CB}^\pp\ \alpha^3 + \epsi_{\BB}^\pp\ \alpha^3$. 
Instead of the exchange of two Breit-photons, Sucher reported the exchange of two transverse (retarded) photons, Eq.~(\ref{eq:ttpp}), and
$(\epsi_{\CB}^\pp + \epsi_{\TT}^\pp)\ \alpha^3$, Eq.~(\ref{eq:cbtt}), 
which is in a reasonable agreement with our numerical result for the non-retarded value, Eq.~(\ref{eq:tbthree}), but the deviation is non-negligible.

We note that the excellent agreement of the variational and corresponding perturbative energies is observed only for the no-pair Hamiltonian, as defined in Secs.~\ref{sec:intro} and \ref{sec:method} with the projector of the non-interacting electrons in the field of the fixed nuclei. 
Regarding the `bare' (unprojected) DC(B) operators,  
or the positive-energy projected DC(B) operator with different projectors (free-particle or modified $Z$ values), 
none of them resulted in a good numerical agreement with the well-established perturbative expressions 
of the `non-relativistic' QED operators.

Afterall, this numerical observation is not so surprising, if we consider the emergence of the no-pair Dirac--Coulomb--Breit operator (with unretarded electron-electron interaction) from the Bethe--Salpeter QED wave equation following Salpeter's \cite{Sa52} and Sucher's \cite{sucherPhD1958}  calculation.
In this context, a historical note about the Breit equation \cite{Breit1929} (eigenvalue equation for DCB without positive-energy projection, `bare' DCB) may also be relevant as it was pointed out by Douglas and Kroll \cite{DoKr74}.  When Breit used his equation in a perturbative procedure, he had to omit an `extra' term to have good agreement with experiment. This erroneous term was shown by Brown and Ravenhall \cite{BrRa51} to correspond to a contribution from negative-energy intermediate states, which, according to Dirac's hole theory, had to be discarded.

\begin{table}
  \caption{%
    Leading-order ($\alpha^2\Eh$) relativistic corrections from the variational procedure without regularization: 
      $\alpha^2 E_\text{DCB}^{(2)}$: leading-order DCB energy obtained by direct evaluation; 
      %and 
      $O_{\alpha^2}[E_\DCpB^\pp]$ and $O_{\alpha^2}[E_\DCB^\pp]$: 
      $\alpha^2\Eh$-order fitted terms to the $\DCpB$ and $\DCB$ energies.
      Deviations are listed, in n$\Eh$, 
      from the regularized reference value, $\alpha^2 E^{(2)}_{\text{DCB},\reg}$ (Table~S13). 
      \label{tab:alpha2DCB} 
    }
    \begin{tabular}{@{}r@{\ \ \ } r d{1.1}d{4.1}d{4.1} d{3.1}d{3.0} d{2.1}d{2.1}d{3.1} @{}}
    \hline\hline\\[-0.35cm]
    \multicolumn{1}{l}{} &
    \multicolumn{1}{c}{H$^-$}   &
    \multicolumn{1}{c}{He ($2S)$} &
    \multicolumn{1}{c}{He ($1S)$} &
    \multicolumn{1}{c}{Li$^+$} &
    \multicolumn{1}{c}{Be$^{2+}$} &
    \multicolumn{1}{c}{H$_2$} &
    \multicolumn{1}{c}{H$_3^+$} &
    \multicolumn{1}{c}{HeH$^+$} 
    \\
    \hline\\[-0.35cm]
    $\alpha^2 E_\text{DCB}^{(2)}- \alpha^2 E_{\text{DCB},\reg}^{(2)}$
    &	1.7	& 21	&-18	&	296	&	766	&	9.0	&	9.3	&	55	\\
    $O_{\alpha^2}[E_\DCpB^\pp]-\alpha^2 E_{\text{DCB},\reg}^{(2)}$
    &	0.4	& 0.5	&4.0	&	13	&	31	&	1.1	&	1.6	&	1.1 \\
    $O_{\alpha^2}[E_\DCB^\pp]-\alpha^2 E_{\text{DCB},\reg}^{(2)}$
    &	0.5	& 0.4	&3.2	&	13	&	19	&	1.1	&	1.6	&	0.6 \\
    \hline\hline
  \end{tabular}
\end{table}

\vspace{0.25cm}
\subsection{Leading-order, $\alpha^2\Eh$, relativistic corrections without regularization}
The $\alpha^2\Eh$-order term obtained from the DC computation with an ECG spatial basis set (optimized for the non-relativistic energy to a ppb relative precision) reproduced the regularized, perturbative benchmark DC energy to a ppb precision \cite{JeFeMa22}. At the same time,
the error of the perturbative DC energy by direct integration 
in the ECG basis was an order of magnitude larger \cite{JeFeMa22,JeIrFeMa22}.

In the present work, we observe a similar improvement for the $\alpha^2\Eh$-order contribution obtained from the $\alpha$ polynomial fit to the variational $\edcpb$ and $\edcb$ energies in comparison with the perturbative DCB energy (expectation value of the Breit--Pauli Hamiltonian).

This behaviour is highlighted in Table~\ref{tab:alpha2DCB} (see also Tables~S9--S12), in which the $\alpha^2\Eh$ energies obtained from the $\alpha$-scaling approach are compared with perturbative values obtained by direct or regularized integration.
The improvement can be explained by the fact that the `singular' operators of the Breit--Pauli Hamiltonian are implicitly included in the sixteen-component Dirac--Coulomb--Breit operator, for which the eigenvalue equation is solved variationally, \emph{i.e.,} the linear combination coefficients of the kinetically balanced ECG basis set are relaxed in a variational manner. Thereby, the relativistic corrections are not \emph{a posteriori} computed as expectation values, but they are automatically included in the variational energy computation.

To generate Figs.~\ref{fig:BreitPTalphafit} and \ref{fig:DCBalphafit} and the fitted $\alpha$ polynomials,
we used the `own basis' value (direct evaluation) of $\langle \delta(\br_{12})\rangle_\text{nr}$ and $\langle H_\text{B}^{(2)}\rangle_\text{nr}$ (Table~S14).
Then, using the fitted coefficients and these two integral values,  
we obtained the leading-order ($\alpha^2\Eh$) relativistic correction (Table~\ref{tab:alpha2DCB}) `carried by' the sixteen-component DC(B) energy.

All in all, the $\alpha^2\Eh$-order 
contribution to the DCB energy is an order of magnitude more accurate than the perturbative correction by direct integration in comparison with the benchmark,  regularized value.

\vspace{0.25cm}
\subsection{Higher-order ($\alpha^4\Eh$) relativistic corrections}
The $\alpha^4\Eh$-order contribution to the DC(B) energy increases with an increasing $Z$ value.
Based on the $\alpha$-scaling plots, we can observe that the ratio of the $\alpha^4\Eh$ to $\alpha^3\Eh$ contribution is 
5 and 10~\% for $Z=2$ (ground state) for the DCB and the DC energy, respectively, but this ratio is already 40 and 50~\% for $Z=4$ (ground state).
Hence, `resummation' in $Z\alpha$ of the perturbative series appears 
to be important for the total energies, $\edc, \edcpb, \edcb$ (Table~\ref{tab:orders}) already for intermediate $Z$ values. 

Regarding the Breit term, the first-order Breit correction to the Coulomb interaction has important contribution at orders $\alpha^2\Eh$ and $\alpha^3\Eh$, but the $\alpha^4\Eh$-order contribution to the no-pair energy remains relatively small (for the systems studied in this work).

{%
The only outlier from these observations is He (2S). The large $a_4\alpha/a_3$ ratio for the DC (and similarly for the $\DCpB$ and DCB) energy can be understood by noting that the $\langle \delta(\bos{r}_{12})\rangle_\text{nr}$ factor (Table~S13), and thus, the thrid-order correction, is very small. 
The third-order DC energy contribution (known to be proportional to $\langle \delta(\bos{r}_{12})\rangle_\text{nr}$ based on perturbation theory, Eq.~(\ref{eq:ccpp})) is $-11$ and $-135$~$\Eh$ for He (2S) and (1S), respectively, whereas the fourth-order terms are comparable,  $-12$~$\Eh$ for both He (2S) and (1S).
}

\begin{table}
  \caption{%
    Relative importance of the $\alpha^4\Eh$- and $\alpha^3\Eh$-order contributions to the no-pair Dirac--Coulomb, Breit, and Dirac--Coulomb--Breit energies.
    The energy contribution is to be understood as $(a_3\alpha^3 + a_4\alpha^4)\langle \delta(\bos{r}_{12}) \rangle_\nonrel$. 
    \label{tab:orders}
  }
    \centering
\scalebox{1.}{%    
    \begin{tabular}{@{}l d{2.2}d{5.2}d{1.0}c d{2.2}d{3.1}d{1.0}c 
    %d{2.2}d{5.2}d{1.0}c 
    d{2.2}d{5.2}d{1.0}c d{2.2}d{5.2}d{1.0} @{}}
    \hline\hline\\[-0.35cm]
	&	
	\multicolumn{3}{c}{$\edc(\alpha)$} & &
	\multicolumn{3}{c}{$\langle \text{B}\rangle_\DC(\alpha)$} & &
	\multicolumn{3}{c}{$\edcpb(\alpha)$} & &
	\multicolumn{3}{c}{$\edcb(\alpha)$} \\
	&	
	\multicolumn{3}{c}{Ref.~\cite{JeFeMa22}} & &
	\multicolumn{3}{c}{Fig.~\ref{fig:BreitPTalphafit}} & &
	\multicolumn{3}{c}{} & &
	\multicolumn{3}{c}{} \\
	\cline{2-4}
	\cline{6-8}
	\cline{10-12}
	\cline{14-16}
	\\[-0.35cm]
	&	
	\multicolumn{1}{c}{$a_3$}	&	\multicolumn{1}{c}{$a_4$}	&	\multicolumn{1}{c}{$\frac{a_4\alpha}{a_3}$~\%}	& &
	\multicolumn{1}{c}{$a_3$}	&	\multicolumn{1}{c}{$a_4$}	&	\multicolumn{1}{c}{$\frac{a_4\alpha}{a_3}$~\%}	& &	
	\multicolumn{1}{c}{$a_3$}	&	\multicolumn{1}{c}{$a_4$}	&	\multicolumn{1}{c}{$\frac{a_4\alpha}{a_3}$~\%}	& &	
	\multicolumn{1}{c}{$a_3$}	&	\multicolumn{1}{c}{$a_4$}	&	\multicolumn{1}{c}{$\frac{a_4\alpha}{a_3}$~\%}	\\
    \hline 
    H$_2$	&	
    -3.27	&	-6.59	&	1	& &	
    10.04	&	2.6	    &	0	& &	
    6.77	&	-3.99	&	0	& &	
    4.60	&	8.41	&	1	\\
    H$_3^+$	&	
    -3.27	&	-5.27	&	1	& &	
    10.13	&	-2.6	&	0	& &	
    6.86	&	-7.87	&	-1	& &	
    4.67	&	4.73	&	1	\\
    HeH$^+$	&	
    -3.27	&	-44.5	&	10	& &	
    10.16	&	-1.3	&	0	& &	
    6.89	&	-45.8	&	-5	& &	
    4.73	&	-39.7	&	-6	\\
    \hline
    H$^-$	&	
    -3.26	&	-23.6	&	5	& &	
    10.05	&	5.7	    &	0	& &	
    6.79	&	-17.9	&	-2	& &	
    4.62	&	-1.6	&	0	\\
    He (2S)	&	
    -3.26	&	-470	& 105 & &	 %DC a3, a4
    10.13	&	1.2 	&	0	& &	 %<B> a3, a4
    6.87	&	-469	& -50	& &	 %DC<B> a3, a4
    4.71 	&	-473	& -73	\\   %DCB a3, a4    
    He (1S)	&	
    -3.27	&	-40.3	&	9	& &	
    10.18	&	1.9 	&	0	& &	
    6.91	&	-38.4	&	-4	& &	
    4.75	&	-35.3	&	-5	\\
    Li$^+$	&	
    -3.26	&	-109	&	24	& &
    10.22	&	4.3	    &	0	& &	
    6.96	&	-105	&	-11	& &	
    4.80	&	-112	&	-17	\\
    Be$^{2+}$	&	
    -3.25	&	-241	&	54	& &	
    10.22	&	11  	&	1	& &	
    6.97	&	-230	&	-24	& &	
    4.80	&	-246	&	-37	\\
    \hline\hline 
    \end{tabular}
}
\end{table}

\clearpage
\section{Summary and conclusion}
Variational and perturbative relativistic energies are computed and compared for two-electron atoms and molecules with low nuclear charge numbers.
In general, good agreement of the two approaches is observed. 
Remaining deviations can be attributed to higher-order relativistic, also called non-radiative quantum electrodynamics (QED), corrections of the perturbative approach that are automatically included in the variational solution of the no-pair Dirac--Coulomb--Breit (DCB) equation to all orders of the $\alpha$ fine-structure constant. 
The analysis of the polynomial $\alpha$ dependence of the DCB energy makes it possible to determine the leading-order relativistic correction to the non-relativistic energy to high precision without regularization. 
Contributions from the Breit--Pauli Hamiltonian, for which expectation values converge slowly due the singular terms, are implicitly included in the variational procedure.
The $\alpha$ dependence of the no-pair DCB energy shows that  
the higher-order ($\alpha^4\Eh$) non-radiative QED correction is 5~\% of the leading-order ($\alpha^3\Eh$) non-radiative QED correction for $Z=2$ (He), but it is 40~\% already for $Z=4$ (Be$^{2+}$), which indicates that resummation provided by the variational procedure is important already for intermediate nuclear charge numbers.

\vspace{0.5cm}
\section*{Data availability statement}
The data that support findings of this study is included in the paper or in the Supplementary Material.

\vspace{0.5cm}
\section*{Supplementary Material}
The supplementary material contains (1) Convergence tables; (2) Leading-order corrections from $\alpha$ scaling; (3) Non-relativistic energies and perturbative corrections.

\vspace{0.5cm}
\begin{acknowledgments}
\noindent Financial support of the European Research Council through a Starting Grant (No.~851421) is gratefully acknowledged. DF thanks a doctoral scholarship from the ÚNKP-21-3 New National Excellence Program of the Ministry for Innovation and Technology from the source of the National Research, Development and Innovation Fund (ÚNKP-21-3-II-ELTE-41). \\
\end{acknowledgments}

%\clearpage
%\bibliography{references}

\clearpage

\setcounter{section}{0}
\renewcommand{\thesection}{S\arabic{section}}
\setcounter{subsection}{0}
\renewcommand{\thesubsection}{S\arabic{section}.\arabic{subsection}}

\setcounter{equation}{0}
\renewcommand{\theequation}{S\arabic{equation}}

\setcounter{table}{0}
\renewcommand{\thetable}{S\arabic{table}}

\setcounter{figure}{0}
\renewcommand{\thefigure}{S\arabic{figure}}

~\\[0.cm]
\begin{center}
\begin{minipage}{0.8\linewidth}
\centering
\textbf{Supplementary Material} \\[0.25cm]

\textbf{Variational versus perturbative relativistic energies for small and light atomic and molecular systems}
\end{minipage}
~\\[0.5cm]
\begin{minipage}{0.6\linewidth}
\centering

D\'avid Ferenc,$^1$ P\'eter Jeszenszki,$^1$ and Edit M\'atyus$^{1,\ast}$ \\[0.15cm]

$^1$~\emph{ELTE, Eötvös Loránd University, Institute of Chemistry, 
Pázmány Péter sétány 1/A, Budapest, H-1117, Hungary} \\[0.15cm]
$^\ast$ edit.matyus@ttk.elte.hu \\
\end{minipage}
~\\[0.15cm]
(Dated: December 3, 2021)
\end{center}

~\\[1cm]
\begin{center}
\begin{minipage}{0.9\linewidth}
\noindent %
Contents: \\
S1. Convergence tables \\
S2. Leading-order corrections from $\alpha$ scaling \\
S3. Non-relativistic energies and perturbative corrections \\
\end{minipage}
\end{center}

\clearpage
%\section{Matrix elements of the Breit operator}
\section{Convergence tables}
  \begin{table}[h]
  \caption{%
  H$^-$ ground state (1 $^1$S$_0$): %  
  Convergence of the energy, in $\Eh$, with respect to the $N_\text{b}$ number of explicitly correlated Gaussian (ECG) basis functions.
  The non-relativistic (nr), Dirac--Coulomb projected ($\edc$), Dirac--Coulomb--Breit projected ($\edcb$), and $\edc$ with Breit perturbative correction ($\langle B\rangle_\DC$, $\mathcal{P}^{(2)}_\DC\lbrace B\rbrace$) energies are shown.
  Double precision arithmetic and a penalty value (that controls the overlap of the generated ECG functions) of 0.99 was used.
  \label{tab:hm}
  }
  \begin{tabular}{@{}l rrr rr @{}}
    \hline\hline\\[-0.40cm]
    \multicolumn{1}{l}{$N_\mathrm{b}$} &
    \multicolumn{1}{c}{$E_\nonrel$ $^\text{a}$}  &
    \multicolumn{1}{c}{$E_\DC^\pp$}  &
    \multicolumn{1}{c}{$E_\DC^\pp+\langle B \rangle_\DC$} &
    \multicolumn{1}{c}{$E_\DC^\pp+\mathcal{P}^{(2)}_\DC\lbrace B\rbrace$} &
    \multicolumn{1}{c}{$E_\DCB^\pp$}
    \\
    \hline\\[-0.35cm]
    300	& $-$0.527 750 974 1  & $-$0.527 756 691 0  & $-$0.527 756 233 3 & $-$0.527 756 234 6 & $-$0.527 756 234 6 \\
    400	& $-$0.527 751 015 5  & $-$0.527 756 732 1  & $-$0.527 756 277 4 & $-$0.527 756 279 6 & $-$0.527 756 279 6 \\
    500	& $-$0.527 751 016 4  & $-$0.527 756 733 0  & $-$0.527 756 278 5 & $-$0.527 756 280 1 & $-$0.527 756 280 7 \\
    600	& $-$0.527 751 016 4  & $-$0.527 756 733 0  & $-$0.527 756 278 6 & $-$0.527 756 280 8 & $-$0.527 756 280 7 \\
    \hline\\[-0.35cm]
    \multicolumn{2}{l}{$E-E^{(2)}_{\text{DCB}}$ $^\text{b}$}  
      &   & \ 0.000 000 074 0  &  0.000 000 005 2  &  0.000 000 005 3  \\
    \multicolumn{2}{l}{$E-E^{(3)}_{\text{DC}\langle\text{B}\rangle}$ $^\text{c}$} 
      &   & $-$0.000 000 000 1  &  $-$0.000 000 002 3 &  $-$0.000 000 002 2 \\
    \multicolumn{2}{l}{$E-E^{(3)}_\text{DCB}$ $^\text{d}$}   
      &    & & $-$0.000 000 000 6 & $-$0.000 000 000 5 \\
    \hline\hline
  \end{tabular}
  \begin{flushleft}
    {\footnotesize
    $^\text{a}$ %    
    $E_\nonrel=-$0.527\ 751\ 016\ 5~$\Eh$\ \cite{Dr06}.}\\
    {\footnotesize    
    $^\text{b}$ %
    $E^{(2)}_\text{DCB}=-$0.527 756 286 0~$\Eh$\ \cite{Dr06}.}\\
    {\footnotesize
    $^\text{c}$ %
    $E^{(3)}_{\text{DC}\langle\text{B}\rangle}=-$0.527 756 278 5~$\Eh$\ .} \\
    {\footnotesize
    $^\text{d}$ %
    $E^{(3)}_\text{DCB}=-$0.527 756 280 2~$\Eh$\ .} 
  \end{flushleft}  
\end{table}

\begin{table}[h]
  \caption{%
    He (2~$^1$S$_0$, $2S$):
    Convergence of the energy, in $\Eh$, with respect to the $N_\text{b}$ number of explicitly correlated Gaussian (ECG) basis functions.
    The non-relativistic (nr), Dirac--Coulomb projected ($\edc$), Dirac--Coulomb--Breit projected ($\edcb$), and $\edc$ with Breit perturbative correction ($\langle B\rangle_\DC$, $\mathcal{P}^{(2)}_\DC\lbrace B\rbrace$) energies are shown.
    Double precision arithmetic and a penalty value (that controls the overlap of the generated ECG functions) of 0.99 was used.
    \label{tab:he2s}}
  \begin{tabular}{@{}l rrr rr@{}}
    \hline\hline\\[-0.40cm]
    \multicolumn{1}{l}{$\nb$} &	
    \multicolumn{1}{c}{$E_\nonrel$ $^\text{a}$}  &
    \multicolumn{1}{c}{$E_\DC^\pp$}  &
    \multicolumn{1}{c}{$E_\DC^\pp+\langle B \rangle_\DC$} &
    \multicolumn{1}{c}{$E_\DC^\pp+\mathcal{P}^{(2)}_\DC\lbrace B\rbrace$} &
    \multicolumn{1}{c}{$E_\DCB^\pp$} 
    \\
    \hline\\[-0.35cm]
    100	& $-$2.145 973 294 &   $-$2.146 084 035	&	$-$2.146 081 584	&	$-$2.146 081 588	&	$-$2.146 081 588	\\
    200	& $-$2.145 974 011 & 	$-$2.146 084 756	&	$-$2.146 082 320	&	$-$2.146 082 327	&	$-$2.146 082 327	\\
    300	& $-$2.145 974 045 & 	$-$2.146 084 789	&	$-$2.146 082 353	&	$-$2.146 082 361	&	$-$2.146 082 361	\\
    400	& $-$2.145 974 046 & 	$-$2.146 084 791	&	$-$2.146 082 355	&	$-$2.146 082 363	&	$-$2.146 082 363	\\
    \hline\\[-0.35cm] 
    $E-E^{(2)}_\text{DCB}$ $^\text{b}$  &  	& 
     &	0.000 000 013	&	0.000 000 005	&	0.000 000 005	\\
    $E-E^{(3)}_{\text{DC}\langle\text{B}\rangle}$ $^\text{c}$  &  	& 
     &	$-$0.000 000 011	&	$-$0.000 000 019	&	$-$0.000 000 019	\\
    $E-E^{(3)}_{\text{DCB}}$ $^\text{d}$  &  	& 
     &		& $-$0.000 000 013	& $-$0.000 000 013	\\
    \hline\hline
    \end{tabular}
    \begin{flushleft}
    {\footnotesize
    $^\text{a}$ %    
    %$E_\nonrel=-$2.145 974 046 1~$\Eh$\ \cite{Dr06}.}\\
    $E_\nonrel=-$2.145 974 046~$\Eh$\ \cite{Dr06}.}\\
    {\footnotesize
    $^\text{b}$ %
    $E^{(2)}_\text{DCB}=-$2.146 082 368~$\Eh$\ \cite{Dr88}.} \\
    {\footnotesize
    $^\text{c}$ %
    $E^{(3)}_{\text{DC}\langle\text{B}\rangle}=
    -2.146\ 082\ 344\ \Eh$.} \\
    {\footnotesize
    $^\text{d}$ %
    $E^{(3)}_\text{DCB}=-2.146\ 082\ 350$~$\Eh$.} 
  \end{flushleft}  
\end{table}

\begin{table}[h]
  \caption{%
  He (1~$^1$S$_0$, $1S$): %
  Convergence of the energy, in $\Eh$, with respect to the $N_\text{b}$ number of explicitly correlated Gaussian (ECG) basis functions.
  The non-relativistic (nr), Dirac--Coulomb projected ($\edc$), Dirac--Coulomb--Breit projected ($\edcb$), and $\edc$ with Breit perturbative correction ($\langle B\rangle_\DC$, $\mathcal{P}^{(2)}_\DC\lbrace B\rbrace$) energies are shown.
  Double precision arithmetic and a penalty value (that controls the overlap of the generated ECG functions) of 0.99 was used.
  \label{tab:he1s}
  }
  \begin{tabular}{@{}l rrr rr @{}}
    \hline\hline\\[-0.40cm]
    \multicolumn{1}{l}{$N_\mathrm{b}$} &
    \multicolumn{1}{c}{$E_\nonrel$ $^\text{a}$}  &
    \multicolumn{1}{c}{$E_\DC^\pp$}  &
    \multicolumn{1}{c}{$E_\DC^\pp+\langle B \rangle_\DC$} &
    \multicolumn{1}{c}{$E_\DC^\pp+\mathcal{P}^{(2)}_\DC\lbrace B\rbrace$} &
    \multicolumn{1}{c}{$E_\DCB^\pp$}
    \\
    \hline\\[-0.35cm]
    200	&	$-$2.903 724 363 9  & $-$2.903 856 618 3  & $-$2.903 828 014 6 & $-$2.903 828 103 4 & $-$2.903 828 103 0 \\
    300	&	$-$2.903 724 375 0	& $-$2.903 856 629 8  & $-$2.903 828 029 3 & $-$2.903 828 118 7 & $-$2.903 828 118 3 \\
    400	&	$-$2.903 724 376 3	& $-$2.903 856 631 3  & $-$2.903 828 031 5 & $-$2.903 828 121 2 & $-$2.903 828 120 7 \\
    500	&	$-$2.903 724 376 6  & $-$2.903 856 631 5  & $-$2.903 828 031 9 & $-$2.903 828 121 5 & $-$2.903 828 121 1 \\
    700	&	$-$2.903 724 376 6	& $-$2.903 856 631 6  & $-$2.903 828 031 9 & $-$2.903 828 121 5 & $-$2.903 828 121 1 \\
    \hline\\[-0.35cm]
    \multicolumn{2}{l}{$E-E^{(2)}_{\text{DCB}}$ $^\text{b}$}
      &   & 0.000 000 278 7  &  0.000 000 189 1  &  0.000 000 189 5 \\
    \multicolumn{2}{l}{$E-E^{(3)}_{\text{DC}\langle\text{B}\rangle}$ $^\text{c}$}
      &   & $-$0.000 000 012 5 &  $-$0.000 000 102 0 &  $-$0.000 000 101 7 \\
    \multicolumn{2}{l}{$E-E^{(3)}_\text{DCB}$ $^\text{d}$}  
      &   & & $-$0.000 000 037 2 & $-$0.000 000 036 8 \\
    \hline\hline
  \end{tabular}
    
  \begin{flushleft}
    {\footnotesize
    $^\text{a}$ %    
    $E_\nonrel=-$2.903 724 377 0~$\Eh$\ \cite{Dr06}.}\\
    {\footnotesize
    $^\text{b}$ %
    $E^{(2)}_\text{DCB}=-$2.903 828 310 6~$\Eh$\ \cite{Dr06}. }\\
    {\footnotesize
    $^\text{c}$ %
    $E^{(3)}_{\text{DC}\langle\text{B}\rangle}=-$2.903 828 019 5~$\Eh$\ .} \\
    {\footnotesize
    $^\text{d}$ %
    $E^{(3)}_\text{DCB}=-$2.903 828 084 4~$\Eh$\ .} 
  \end{flushleft}  
  \end{table}

\begin{table}[h]
  \caption{%
    Li$^+$ (1~$^1\text{S}_0$):
    Convergence of the energy, in $\Eh$, with respect to the $N_\text{b}$ number of explicitly correlated Gaussian (ECG) basis functions.
    The non-relativistic (nr), Dirac--Coulomb projected ($\edc$), Dirac--Coulomb--Breit projected ($\edcb$), and $\edc$ with Breit perturbative correction ($\langle B\rangle_\DC$, $\mathcal{P}^{(2)}_\DC\lbrace B\rbrace$) energies are shown.
    \label{tab:Lip}
  }
    \begin{tabular}{@{}l rrrrr @{}}
    \hline\hline\\[-0.40cm]
    \multicolumn{1}{l}{$N_\mathrm{b}$ $^\text{a}$} & 
    \multicolumn{1}{c}{$E_\nonrel$}  &
    \multicolumn{1}{c}{$E_\DC^\pp$}  &
    \multicolumn{1}{c}{$E_\DC^\pp+\langle B \rangle_\DC$} &
    \multicolumn{1}{c}{$E_\DC^\pp+\mathcal{P}^{(2)}_\DC\lbrace B\rbrace$} &  
    \multicolumn{1}{c}{$E_\DCB^\pp$}  
    \\
    \hline\\[-0.35cm]
    100	& $-$7.279 913 057 & $-$7.280 698 534 & $-$7.280 540 513 & $-$7.280 540 955 & $-$7.280 540 953 \\
    200	& $-$7.279 913 380 & $-$7.280 698 868 & $-$7.280 540 943 & $-$7.280 541 409 & $-$7.280 541 407 \\
    300	& $-$7.279 913 409 & $-$7.280 698 887 & $-$7.280 540 964 & $-$7.280 541 431 & $-$7.280 541 429 \\
    400	& $-$7.279 913 410 & $-$7.280 698 899 & $-$7.280 540 978 & $-$7.280 541 445 & $-$7.280 541 443 \\
    \hline\\[-0.35cm]
    \multicolumn{2}{l}{$E-E^{(2)}_{\text{DCB}}$ $^\text{b}$}
      &   &    0.000 001 300  &   0.000 000 833  &  0.000 000 835 \\
    \multicolumn{2}{l}{$E-E^{(3)}_{\text{DC}\langle\text{B}\rangle}$ $^\text{c}$}
      &   & $-$0.000 000 161  &  $-$0.000 000 628  &  $-$0.000 000 626 \\
    \multicolumn{2}{l}{$E-E^{(3)}_\text{DCB}$ $^\text{d}$}  
      &   &  &  $-$0.000 000 303  &  $-$0.000 000 301  \\
    \hline\hline
    \end{tabular}
    \begin{flushleft}
    {\footnotesize
    $^\text{a}$ %    
    $E_\nonrel=-$7.279 913 413~$\Eh$\ \cite{Dr06}.} \\
    {\footnotesize
    $^\text{b}$ %
    $E^{(2)}_\text{DCB}=-$7.280 542 278~$\Eh$\ \cite{Dr06}. }\\
    {\footnotesize
    $^\text{c}$ %
    $E^{(3)}_{\text{DC}\langle\text{B}\rangle}=-$7.280 540 817~$\Eh$\ .} \\
    {\footnotesize
    $^\text{d}$ %
    $E^{(3)}_\text{DCB}=-$7.280 541 143~$\Eh$\ .} 
    \end{flushleft}
\end{table}

\begin{table}[h]
  \caption{%
    Be$^{2+}$ (1~$^1\text{S}_0$): %
    Convergence of the energy, in $\Eh$, with respect to the $N_\text{b}$ number of explicitly correlated Gaussian (ECG) basis functions.
    The non-relativistic (nr), Dirac--Coulomb projected ($\edc$), Dirac--Coulomb--Breit projected ($\edcb$), and $\edc$ with Breit perturbative correction ($\langle B\rangle_\DC$, $\mathcal{P}^{(2)}_\DC\lbrace B\rbrace$) energies are shown.
    \label{tab:Bepp}
  }
  \begin{tabular}{@{}l rrrrr @{}}
    \hline\hline\\[-0.40cm]
    \multicolumn{1}{l}{$N_\mathrm{b}$} &
    \multicolumn{1}{c}{$E_\nonrel$ $^\text{a}$}  &
    \multicolumn{1}{c}{$E_\DC^\pp$}  &
    \multicolumn{1}{c}{$E_\DC^\pp+\langle B \rangle_\DC$} &
    \multicolumn{1}{c}{$E_\DC^\pp+\mathcal{P}^{(2)}_\DC\lbrace B\rbrace$} &  
    \multicolumn{1}{c}{$E_\DCB^\pp$}  
    \\
    \hline\\[-0.35cm]
    100	& $-$13.655 565 910  & $-$13.658 257 248 & $-$13.657 788 246 & $-$13.657 789 601 & $-$13.657 789 595  \\
    200	& $-$13.655 566 229  & $-$13.658 257 596 & $-$13.657 788 720 & $-$13.657 790 097 & $-$13.657 790 090 \\
    300	& $-$13.655 566 234  & $-$13.658 257 602 & $-$13.657 788 729 & $-$13.657 790 107 & $-$13.657 790 100 \\
    \hline\\[-0.35cm]
    \multicolumn{2}{l}{$E-E^{(2)}_{\text{DCB}}$ $^\text{b}$}
      &   & 0.000 003 175 & 0.000 001 797 & 0.000 001 804   \\
    \multicolumn{2}{l}{$E-E^{(3)}_{\text{DC}\langle\text{B}\rangle}$ $^\text{c}$}
      &   & $-$0.000 000 995 & $-$0.000 002 373   &  $-$0.000 002 366  \\
    \multicolumn{2}{l}{$E-E^{(3)}_\text{DCB}$ $^\text{d}$}  
      &   &   & $-$0.000 001 443   &  $-$0.000 001 436  \\
    \hline\hline      
  \end{tabular}
  \begin{flushleft}
    {\footnotesize
    $^\text{a}$ %    
    $E_\nonrel=-$13.655 566 238~$\Eh$\ \cite{Dr06}.} \\
    {\footnotesize
    $^\text{b}$ %
    $E^{(2)}_\text{DCB}=-$13.657 791 904~$\Eh$\ \cite{Dr06}. }\\
    {\footnotesize
    $^\text{c}$ %
    $E^{(3)}_{\text{DC}\langle\text{B}\rangle}=-$13.657 787 730~$\Eh$.} \\
    {\footnotesize
    $^\text{d}$ %
    $E^{(3)}_\text{DCB}=-$13.657 788 664~$\Eh$.} 
  \end{flushleft}
\end{table}
  
\begin{table}[h!]
  \caption{%
    H$_2$ molecule ($X\ ^1\Sigma_\mathrm{g}^+$, $R_\text{pp}=1.4$~bohr):
    Convergence of the energy, in $\Eh$, with respect to the $N_\text{b}$ number of explicitly correlated Gaussian (ECG) basis functions.
    The non-relativistic (nr), Dirac--Coulomb projected ($\edc$), Dirac--Coulomb--Breit projected ($\edcb$), and $\edc$ with Breit perturbative correction ($\langle B\rangle_\DC$, $\mathcal{P}^{(2)}_\DC\lbrace B\rbrace$) energies are shown.
    \label{tab:h2}
  }
  \begin{tabular}{@{}l rrr rr@{}}
    \hline\hline\\[-0.40cm]
    \multicolumn{1}{c}{$\nb$} &	
    \multicolumn{1}{c}{$E_\nonrel$ $^\text{a}$}  &    
    \multicolumn{1}{c}{$E_\DC^\pp$}  &
    \multicolumn{1}{c}{$E_\DC^\pp+\langle B \rangle_\DC$} &
    \multicolumn{1}{c}{$E_\DC^\pp+\mathcal{P}^{(2)}_\DC\lbrace B\rbrace$} &
    \multicolumn{1}{c}{$E_\DCB^\pp$} \\
    \hline\\[-0.35cm]
    128	& $-$1.174 475 542 7 &	$-$1.174 489 583  	&	$-$1.174 486 441  	&	$-$1.174 486 451  	&	$-$1.174 486 451	\\
    256	& $-$1.174 475 697 9 &	$-$1.174 489 738  	&	$-$1.174 486 604  	&	$-$1.174 486 617  	& $-$1.174 486 617	\\
    512	& $-$1.174 475 712 8 &	$-$1.174 489 753  	&	$-$1.174 486 620  	&	$-$1.174 486 633  	&	$-$1.174 486 633	\\
    700	& $-$1.174 475 713 6&	$-$1.174 489 754  	&	$-$1.174 486 621  	&	$-$1.174 486 635  	&	$-$1.174 486 635	\\
    800	& $-$1.174 475 713 6&	$-$1.174 489 754  	&	$-$1.174 486 621  	&	$-$1.174 486 635  	&	$-$1.174 486 635	\\
    1000& $-$1.174 475 713 6&	$-$1.174 489 754  	&	$-$1.174 486 621  	&	$-$1.174 486 635  	&	$-$1.174 486 635	
    \\
    1200& $-$1.174 475 713 9 &	$-$1.174 489 754  	&	$-$1.174 486 622  	&	$-$1.174 486 635  	&	$-$1.174 486 635	\\
    \hline\\[-0.35cm]
    \multicolumn{2}{l}{$E-E^{(2)}_\text{DCB}$ $^\text{b}$}  &  
       	&	0.000 000 045	&	0.000 000 032	&	0.000 000 032	\\
    \multicolumn{2}{l}{$E-E^{(3)}_{\text{DC}\langle\text{B}\rangle}$ $^\text{c}$}  &  
        &	$-$0.000 000 001	&	$-$0.000 000 014	&	$-$0.000 000 014	\\
    \multicolumn{2}{l}{$E-E^{(3)}_\text{DCB}$ $^\text{d}$}  &  
       	&		& $-$0.000 000 004 & $-$0.000 000 004	\\
    \hline\hline
    \end{tabular}
  \begin{flushleft}
    {\footnotesize
    $^\text{a}$ %    
    $E_\nonrel=-$1.174\ 475\ 714\ 2\ $\Eh$\ \cite{PuKoPa17}.} \\
    {\footnotesize
    $^\text{b}$ %
    $E^{(2)}_{\text{DCB}}=-$1.174 486 667~$\Eh$\ \cite{PuKoPa17,PuKoCzPa16}.} \\
    {\footnotesize    
    $^\text{c}$ %
    $E^{(3)}_{\text{DC}\langle\text{B}\rangle}=-$1.174 486 621~$\Eh$.} \\
    {\footnotesize
    $^\text{d}$ %
    $E^{(3)}_{\text{DCB}}=-$1.174 486 631~$\Eh$.}
  \end{flushleft}  
  
  \end{table}
  
  \begin{table}[h!]
  \caption{%
    H$_3^+$ molecular ion (ground state, $R_\text{pp}=1.65$~bohr):
    Convergence of the energy, in $\Eh$, with respect to the $N_\text{b}$ number of explicitly correlated Gaussian (ECG) basis functions.
    The non-relativistic (nr), Dirac--Coulomb projected ($\edc$), Dirac--Coulomb--Breit projected ($\edcb$), and $\edc$ with Breit perturbative correction ($\langle B\rangle_\DC$, $\mathcal{P}^{(2)}_\DC\lbrace B\rbrace$) energies are shown.
    The basis set size corresponds to the use of $D_{3\text{h}}$
    point-group symmetry in the computations.      
    \label{tab:h3p}
  }
  \begin{tabular}{@{}l rrr rr@{}}
    \hline\hline\\[-0.40cm]
    \multicolumn{1}{l}{$\nb$} &	
    \multicolumn{1}{c}{$E_\nonrel$ $^\text{a}$} &    
    \multicolumn{1}{c}{$E_\DC^\pp$} &
    \multicolumn{1}{c}{$E_\DC^\pp+\langle B \rangle_\DC$} &
    \multicolumn{1}{c}{$E_\DC^\pp+\mathcal{P}^{(2)}_\DC\lbrace B\rbrace$} &
    \multicolumn{1}{c}{$E_\DCB^\pp$} \\
    \hline \\[-0.35cm]
    100   & $-$1.343 835 248 6& $-$1.343 850 149 & $-$1.343 846 977 & $-$1.343 846 988 & $-$1.343 846 988 \\
    200   & $-$1.343 835 605 7& $-$1.343 850 507 & $-$1.343 847 343 & $-$1.343 847 357 & $-$1.343 847 357 \\
    300   & $-$1.343 835 623 1& $-$1.343 850 524 & $-$1.343 847 363 & $-$1.343 847 378 & $-$1.343 847 378 \\
    400   & $-$1.343 835 624 9& $-$1.343 850 526 & $-$1.343 847 365 & $-$1.343 847 380 & $-$1.343 847 380 \\
    500   & $-$1.343 835 625 3& $-$1.343 850 527 & $-$1.343 847 366 & $-$1.343 847 381 & $-$1.343 847 381 \\
    600   & $-$1.343 835 625 4& $-$1.343 850 527 & $-$1.343 847 366 & $-$1.343 847 381 & $-$1.343 847 381 \\    
    \hline \\[-0.35cm]
    \multicolumn{2}{l}{$E-E^{(2)}_\text{DCB}$ $^\text{b}$}  &  
    &	 0.000 000 050	&	0.000 000 035	&	0.000 000 035	\\
	\multicolumn{2}{l}{$E-E^{(3)}_{\text{DC}\langle\text{B}\rangle}$ $^\text{c}$}  &  
    &	 0.000 000 000	&	$-$0.000 000 015	&	$-$0.000 000 015	\\
	\multicolumn{2}{l}{$E-E^{(3)}_{\text{DCB}}$ $^\text{d}$}  &  
    &	 	&	$-$0.000 000 004 	&	$-$0.000 000 004	\\    
    \hline\hline
    \end{tabular}
  \begin{flushleft}
    {\footnotesize
    $^\text{a}$ %    
    $E_\nonrel=-$1.343\ 835\ 625\ 4~$\Eh$\ \cite{JeIrFeMa22}.} \\
    {\footnotesize    
    $^\text{b}$ %
    $E^{(2)}_\text{DCB}=-$1.343 847 416~$\Eh$\ \cite{JeIrFeMa22}.} \\
    {\footnotesize
    $^\text{c}$ %
    $E^{(3)}_{\text{DC}\langle\text{B}\rangle}=-$1.343 847 366~$\Eh$.} \\
    {\footnotesize    
    $^\text{d}$ %
    $E^{(3)}_\text{DCB}=-$1.343 847 377~$\Eh$.} 
  \end{flushleft}    
\end{table}

  \begin{table}[h!]
  \caption{%
    HeH$^+$ molecular ion (ground state, $R_{\alpha\text{p}}=1.46$~bohr):
    Convergence of the energy, in $\Eh$, with respect to the $N_\text{b}$ number of explicitly correlated Gaussian (ECG) basis functions.
    The non-relativistic (nr), Dirac--Coulomb projected ($\edc$), Dirac--Coulomb--Breit projected ($\edcb$), and $\edc$ with Breit perturbative correction ($\langle B\rangle_\DC$, $\mathcal{P}^{(2)}_\DC\lbrace B\rbrace$) energies are shown.
    \label{tab:hehp}
  }
  \begin{tabular}{@{}l rrr rr@{}}
    \hline\hline\\[-0.40cm]
    \multicolumn{1}{c}{$\nb$} &	
    \multicolumn{1}{c}{$E_\nonrel$ $^\text{a}$}  &    
    \multicolumn{1}{c}{$E_\DC^\pp$}  &
    \multicolumn{1}{c}{$E_\DC^\pp+\langle B \rangle_\DC$} &
    \multicolumn{1}{c}{$E_\DC^\pp+\mathcal{P}^{(2)}_\DC\lbrace B\rbrace$} &
    \multicolumn{1}{c}{$E_\DCB^\pp$} \\
    \hline\\[-0.35cm]
    400	 & $-$2.978 706 548 8 & $-$2.978 834 584	&	$-$2.978 807 858	&	$-$2.978 807 941	&	$-$2.978 807 941	\\
    600	 &$-$2.978 706 593 3  & $-$2.978 834 630	&	$-$2.978 807 912	&	$-$2.978 807 996	&	$-$2.978 807 996	\\
    800	 & $-$2.978 706 597 3 & $-$2.978 834 634	&	$-$2.978 807 917	&	$-$2.978 808 002	&	$-$2.978 808 001	\\
    1000 & $-$2.978 706 598 2 & $-$2.978 834 635	&	$-$2.978 807 918	&	$-$2.978 808 003	&	$-$2.978 808 003	\\
    1200 &  $-$2.978 706 598 5 & $-$2.978 834 635	&	$-$2.978 807 919	&	$-$2.978 808 004	&	$-$2.978 808 003	\\
    \hline\\[-0.35cm]
    \multicolumn{2}{l}{$E-E^{(2)}_\text{DCB}$ $^\text{b}$}  &  
    &	0.000 000 261	&	0.000 000 176	&	0.000 000 177	\\
    \multicolumn{2}{l}{$E-E^{(3)}_{\text{DC}\langle \text{B}\rangle}$ $^\text{c}$}  &  
    &	$-$0.000 000 016	&	$-$0.000 000 101	&	$-$0.000 000 100	\\
    \multicolumn{2}{l}{$E-E^{(3)}_\text{DCB}$ $^\text{d}$}  &  
    &		&	$-$0.000 000 039	&	$-$0.000 000 038	\\
    \hline\hline
    \end{tabular}
  \begin{flushleft}
    {\footnotesize
    $^\text{a}$ %    
    $E_\nonrel=-$2.978 706 598 5~$\Eh$\ \cite{JeFeMa21,JeFeMa22,FeJeMa22} (\& this work).} \\
    {\footnotesize    
    $^\text{b}$ %
    $E^{(2)}_\text{DCB}=-$2.978 808 180~$\Eh$\ \cite{JeFeMa21}.} \\
    {\footnotesize        
    $^\text{c}$ %
    $E^{(3)}_{\text{DC}\langle\text{B}\rangle}=-$2.978 807 903~$\Eh$.} \\
    {\footnotesize            
    $^\text{d}$ %
    $E^{(3)}_\text{DCB}=-$2.978 807 965~$\Eh$.} 
  \end{flushleft}    
\end{table}

%%%%%%%%%%%%%%%%%%%%%%%%%%%%%%%%%%%%%%%%%%%%%%%%%%%%%%%%%%%%%%%%%%%%%%%
\clearpage
\section{Leading-order corrections from $\alpha$ scaling}
%%%%%%%%%%%%%%%%%%%%%%%%%%%%%%%%%%%%%%%%%%%%%%%%%%%%%%%%%%%%%%%%%%%%%%%
\begin{table}[H]
\caption{%
  Leading-order DC correction in units of $\alpha^2\Eh$.
  $\langle H_\DC^{(2)}\rangle_\nonrel$: direct evaluation of the expectation value of the perturbative DC correction terms, Eq.~(16), in the largest ECG basis set (Tables~\ref{tab:hm}--\ref{tab:hehp});
  $O_{\alpha^2}[E_\DC^\pp]/\alpha^2$: second-order coefficient ($c_2$) of the $\alpha$ dependence of the DC$^\pp$ energy fitted with a third-order polynomial (see also Ref.~\cite{JeFeMa22}).
  $\langle H_\DC^{(2)}\rangle_\reg$: regularized reference values for the perturbative DC corrections are collected (with citations) in Table~\ref{tab:som-pt-lit}.
  \label{tab:alpha2DCcomp}
  }
  \centering
  \begin{tabular}{@{}l r@{\ }r@{\ }r@{\ }r@{\ }r@{\ }r@{\ }r@{\ }r @{}}
    \hline\hline\\[-0.35cm]
    \multicolumn{1}{l}{} &
    \multicolumn{1}{c}{H$^-$}   &
    \multicolumn{1}{c}{He ($2S)$} &
    \multicolumn{1}{c}{He ($1S)$} &
    \multicolumn{1}{c}{Li$^+$} &
    \multicolumn{1}{c}{Be$^{2+}$} &
    \multicolumn{1}{c}{H$_2$} &
    \multicolumn{1}{c}{H$_3^+$} &
    \multicolumn{1}{c}{HeH$^+$} \\
    \hline\\[-0.33cm]
    %$\epsilon_\mathrm{dir}$
    $\langle H_\DC^{(2)}\rangle_\nonrel$
    & $-$0.107 279  & $-$2.078 929 & $-$2.481 823 & $-$14.731 566 & $-$50.477 690 & $-$0.263 240 & $-$0.279 367 & $-$2.401 315 \\
    %
    %$\epsilon_\mathrm{DC}^3$	     
    $O_{\alpha^2}[E_\DC^\pp]/\alpha^2$
    & $-$0.107 279 & $-$2.079 251 & $-$2.480 832 & $-$14.734 771 & $-$50.485 217 & $-$0.263 250 & $-$0.279 386 & $-$2.401 752 \\
    \hline
    %$\epsilon_\mathrm{ref}$
    $\langle H_\DC^{(2)}\rangle_\reg$ (ref.)
    & $-$0.107 283 & $-$2.079 256 & $-$2.480 848 & $-$14.734 859 & $-$50.485 330 & $-$0.263 255 & $-$0.279 399 & $-$2.401 709 \\
    \hline\hline
    \end{tabular}
\end{table}

\begin{table}[h]
  \caption{%
    Deviation, in n$\Eh$, of the leading-order ($\alpha^2\Eh$) DC correction obtained by 
    direct evaluation, 
    $\alpha^2 \langle H_\DC^{(2)}\rangle_\nonrel$, 
    as the second-order term for the $\alpha$ dependence of the $\DC$ energy, $O_{\alpha^2}[E_\DC^\pp]$,
    from the regularized reference value, 
    $\alpha^2 \langle H_\DC^{(2)}\rangle_\reg$, 
    (Table~\ref{tab:som-pt-lit}).
    \label{tab:alpha2DC}
  }    
  \begin{tabular}{@{}l d{2.1}d{2.1}d{3.1}d{4.1} d{4.1}d{2.1}d{2.1}d{2.1} @{}}
    \hline\hline\\[-0.35cm]
    \multicolumn{1}{l}{} &
    \multicolumn{1}{c}{H$^-$}   &
    \multicolumn{1}{c}{He ($2S$)} &    
    \multicolumn{1}{c}{He ($1S$)} &
    \multicolumn{1}{c}{Li$^+$} &
    \multicolumn{1}{c}{Be$^{2+}$} &
    \multicolumn{1}{c}{H$_2$} &
    \multicolumn{1}{c}{H$_3^+$} &
    \multicolumn{1}{c}{HeH$^+$}
    \\
    \hline\\[-0.35cm]
    %$\Delta\epsilon_\mathrm{direct}$         
    $\alpha^2 E_\DC^{(2)}- \alpha^2 \langle H_\DC^{(2)}\rangle_\reg$
    & 0.2 
    &  17.4
    & -52
    & 175
    & 407
    & 0.8
    & 1.7 
    & 21 \\
    %
    %$\Delta\epsilon_\mathrm{DC}^2$       & 0.5 & 18 & 234    & 1450 & 0.7 & 14 & 1.1 \\
    %
    %$\Delta\epsilon_\mathrm{DC}^3$       
    $O_{\alpha^2}[E_\DC^\pp]-\alpha^2 \langle H_\DC^{(2)}\rangle_\reg$
    & 0.2 
    & 0.3
    & 0.9 & 4.7   & 6.0  & 0.3 & 0.7 & -2.3 \\
    \hline\hline
    \end{tabular}
  \end{table}
  
\begin{table}[H]
  \caption{%
  Breit corrections in $\alpha^2\Eh$ units.
  $\langle H_\Breit^{(2)}\rangle_\nonrel$:
  the expectation value (without regularization) of the perturbative correction,  Eq.~(17), in the largest ECG basis set (Tables~\ref{tab:hm}--\ref{tab:hehp});
  $O_{\alpha^2}[\langle B\rangle_\DC]/\alpha^2$ and $O_{\alpha^2}[E_\DCB^\pp-E_\DC^\pp] / \alpha^2$:  
  second-order coefficient of the quartic $\alpha$ polynomial fitted to 
  $\langle B\rangle_\DC$ and $E_\DCB^\pp-E_\DC^\pp$; 
  and
  $\langle H_\Breit^{(2)}\rangle_\reg$:
  regularized reference value of the expectation value of Eq.~(17) with the non-relativistic wave function
  (Table~\ref{tab:som-pt-lit}).
  \label{tab:alpha2Bcomp}
  }
  \begin{tabular}{@{}l rrrrrrrrr @{}}
    \hline\hline\\[-0.35cm]
    \multicolumn{1}{l}{} &
    \multicolumn{1}{c}{H$^-$}   &
    \multicolumn{1}{c}{He $(2S)$} &    
    \multicolumn{1}{c}{He $(1S)$} &
    \multicolumn{1}{c}{Li$^+$} &
    \multicolumn{1}{c}{Be$^{2+}$} &
    \multicolumn{1}{c}{H$_2$} &
    \multicolumn{1}{c}{H$_3^+$} &
    \multicolumn{1}{c}{HeH$^+$} 
    \\
    \hline\\[-0.35cm]
    %$\epsilon_\mathrm{dir}$         
    $\alpha^2 \langle H_\Breit^{(2)}\rangle_\nonrel$
    & 0.008 355 & 0.045 150  & 0.529 738 & 2.927 750 & 8.696 609 & 0.057 721 & 0.058 126 & 0.494 761 \\
    %
    %$\epsilon_{\langle B\rangle}^3$ 
    $O_{\alpha^2}[\langle B\rangle_\DC]/\alpha^2$
    & 0.008 332 & 0.045 095 & 0.529 153 & 2.925 643 & 8.690 333 & 0.057 582 & 0.058 000 & 0.494 192 \\
    %
    %$\epsilon_\mathrm{DCB}^3$	     
    $O_{\alpha^2}[E_\DCB^\pp-E_\DC^\pp] / \alpha^2$
    & 0.008 333 & 0.045 092 & 0.529 138 & 2.925 643 & 8.690 108 & 0.057 582 & 0.058 000 & 0.494 183 \\ \hline
    $\alpha^2 \langle H_\Breit^{(2)}\rangle_\reg$ (ref.)
    & 0.008 328 & 0.045 090 & 0.529 093 & 2.925 486 & 8.689 865 & 0.057 567 & 0.057 983 & 0.494 128 \\
    \hline\hline
    \end{tabular}
\end{table}

\begin{table}[h]
  \caption{%
      Deviation, in n$\Eh$, of 
      the leading-order Breit correction ($\alpha^2\Eh$) obtained by direct (non-regularized) expectation value of the perturbative expression, 
      $\alpha^2 \langle H_\Breit^{(2)}\rangle_\nonrel$, 
      second-order coefficient in the  $\alpha$ polynomial fit (Figs. 1 and 2) of the $\langle B \rangle_\DC$ and of the $E_\DCB^\pp-E_\DC^\pp$ energies
      from the regularized, $\alpha^2\Eh$ order perturbative reference value, 
      $\alpha^2 \langle H_\Breit^{(2)}\rangle_\reg$.
    \label{tab:alpha2Breit}
  }
    \begin{tabular}{@{}l d{2.1}d{2.1}d{3.1}d{4.1} d{4.1}d{3.1}d{3.1}d{2.1} @{}}
    \hline\hline\\[-0.35cm]
    \multicolumn{1}{l}{} &
    \multicolumn{1}{c}{H$^-$}   &
    \multicolumn{1}{c}{He ($2S$)} &    
    \multicolumn{1}{c}{He ($1S$)} &
    \multicolumn{1}{c}{Li$^+$} &
    \multicolumn{1}{c}{Be$^{2+}$} &
    \multicolumn{1}{c}{H$_2$} &
    \multicolumn{1}{c}{H$_3^+$} &
    \multicolumn{1}{c}{HeH$^+$} 
    \\
    \hline\\[-0.35cm]
    %$\Delta\epsilon_\mathrm{dir}$
    $\alpha^2 E_\text{B}^{(2)}-\alpha^2 \langle H_\Breit^{(2)}\rangle_\reg$
    & 1.4 & 3.2  & 34  & 121 & 359 & 8.2 & 7.6 & 33.7 \\
    %
    %$\Delta\epsilon_{\langle B\rangle}^3$ 
    $O_{\alpha^2}[\langle B\rangle_\DC]-\alpha^2 \langle H_\Breit^{(2)}\rangle_\reg$
    & 0.2 & 0.3 & 3.2 & 8.4 &  25 & 0.8 & 0.9 & 3.4 \\    
    %
    %$\Delta\epsilon_\mathrm{DCB}^3$       
    $O_{\alpha^2}[E_\DCB^\pp-E_\DC^\pp]-\alpha^2 \langle H_\Breit^{(2)}\rangle_\reg$
    & 0.3  &  0.1  & 2.4 & 8.4 &  13 & 0.8 & 0.9 & 2.9 \\ 
    \hline\hline
  \end{tabular}
\end{table}

\clearpage
\section{Non-relativistic energies and perturbative corrections}

\begin{table}[H]
  \caption{%
    Reference values for the non-relativistic energy and expectation values for the relativistic correction terms with the non-relativistic wave function. $\langle O\rangle_\reg$ labels regularized evaluation of the expectation value of the singular operators with the non-relativistic wave function. 
    Ground-state values are shown, unless otherwise indicated.
    \label{tab:som-pt-lit}
  }
  \centering
\scalebox{0.93}{%  
  \begin{tabular}{@{}l rr@{\ \ \ } rrr c@{}}
    \hline\hline\\[-0.35cm]
    	&
    \multicolumn{1}{c}{$E_\text{nr}$}	&	
    \multicolumn{1}{c}{$-\frac{1}{8}\sum_{i} \langle (\bos{p}_i^2)^2 \rangle_\reg$}	&	
    \multicolumn{1}{c}{$\sum_{i,a} Z_a \langle \delta(\bos{r}_{ia})\rangle_\reg$}	&	
    \multicolumn{1}{c}{$\langle \delta(\bos{r}_{12})\rangle_\reg$} &
    \multicolumn{1}{c}{$\langle H_\text{OO}\rangle_\nonrel$} & 
    \multicolumn{1}{c}{Ref.} \\
    \hline\\[-0.35cm]
    H$^-$       &	$-$0.527 751 016	&	$-$0.615 640	&	0.329 106	&	0.002 738	&	$-$0.008 875	& \cite{Dr06} \\
    He ($2S$)   &		$-$2.145 974 046	&	$-$10.279 669	    & 5.237 843 &	0.008 648 & $-$0.009 253 & \cite{drakeEnergiesRelativisticCorrections1992} \\    
    He ($1S$)   &	$-$2.903 724 377	&	$-$13.522 017	&	7.241 717	&	0.106 345	&	$-$0.139 095	& \cite{Dr06} \\
    Li$^+$	    &	$-$7.279 913 413	&	$-$77.636 788	&	41.112 057	&	0.533 723	&	$-$0.427 992	& \cite{Dr06} \\
    Be$^{2+}$	&	$-$13.655 566 238	&	$-$261.819 623	&	137.585 380	&	1.522 895	&	$-$0.878 769	& \cite{Dr06} \\
    \hline \\[-0.35cm]													
    H$_2$ $^\text{a}$	&	$-$1.174 475 714	&	$-$1.654 745	&	0.919 336	&	0.016 743	&	$-$0.047 634	&	\cite{PuKoPa17} \\
    H$_3^+$ $^\text{a}$	&	$-$1.343 835 625	&	$-$1.933 424	&	1.089 655	&	0.018 335	&	$-$0.057 218	&	\cite{JeIrFeMa22} \\
    HeH$^+$ $^\text{a}$	&	$-$2.978 706 599	&	$-$13.419 287	&	7.216 253	&	0.101 122	&	$-$0.141 242	& \cite{JeFeMa21,JeFeMa22,FeJeMa22} \\
    \hline\hline\\[-0.35cm]
    	&	
    \multicolumn{1}{c}{$\langle H_\DC^{(2)} \rangle_\reg$} & 
    \multicolumn{1}{c}{$\langle H_\text{B}^{(2)} \rangle_\reg$} & 
    \multicolumn{1}{c}{$\langle H_\DC^{(2)}+H_\text{B}^{(2)} \rangle_\reg$} & 	
    \multicolumn{1}{c}{${E}^{(2)}_\DC$} &	
    \multicolumn{1}{c}{${E}^{(2)}_\DCB$}	&	
    \multicolumn{1}{c}{Ref.}	\\
    \hline\\[-0.35cm]
    H$^-$	   &	$-$0.107 283	&	0.008 328	&	$-$0.098 955	&	$-$0.527 756 730	&	$-$0.527 756 286	& \cite{Dr06}		\\
    He ($2S$)  & $-$2.079 256 & 0.045 090 & $-$2.034 166 & $-$2.146 084 769	&	 $-$2.146 082 368 & \cite{drakeEnergiesRelativisticCorrections1992}  \\    
    He ($1S$)  &	$-$2.480 848	&	0.529 093	&	$-$1.951 755	&	$-$2.903 856 486	&	$-$2.903 828 311	& \cite{Dr06}\\
    Li$^+$	   &	$-$14.734 859	&	2.925 486	&	$-$11.809 373	&	$-$7.280 698 064	&	$-$7.280 542 278	& \cite{Dr06} \\
    Be$^{2+}$  &	$-$50.485 330	&	8.689 865	&	$-$41.795 465	&	$-$13.658 254 651	&	$-$13.657 791 904	& \cite{Dr06}\\
    \hline\\[-0.35cm]
    H$_2$ $^\text{a}$	&  $-$0.263 255	&	0.057 567	&	$-$0.205 689	&	$-$1.174 489 733	&	$-$1.174 486 667	& \cite{PuKoPa17}	\\
    H$_3^+$ $^\text{a}$	&  $-$0.279 399	&	0.057 983	&	$-$0.221 416	&	$-$1.343 850 503	&	$-$1.343 847 416	& \cite{JeIrFeMa22} \\
    HeH$^+$ $^\text{a}$	&  $-$2.401 709	&	0.494 128	&	$-$1.907 581	&	$-$2.978 834 493	&	$-$2.978 808 180	& \cite{JeFeMa21} \\
    \hline\hline\\[-0.35cm]
  \end{tabular}
}
  \begin{flushleft}
    {\footnotesize
    \vspace{-0.20cm}
    $^\text{a}$ $R_\text{pp}=1.4$~bohr, $R_\text{pp}=1.65$~bohr, $R_\text{pp}=1.46$~bohr.}
  \end{flushleft}
\end{table}

\begin{table}[H]
  \caption{%
    Non-relativistic energy and expectation values of the relativistic correction terms obtained in `direct' (non-regularized) computation with the  non-relativistic wave function using the largest basis sets used in this work (Tables~\ref{tab:hm}--\ref{tab:hehp}). In this table, $\langle O\rangle_\text{nr}$ means direct (non-regularized) evaluation of the expectation value with the non-relativistic wave function. 
    Ground-state values are shown, unless otherwise indicated.
    \label{tab:som-pt-own}
  }
  \centering
\scalebox{0.93}{%  
  \begin{tabular}{@{}l rr@{\ \ \ } rrr @{}}
    \hline\hline\\[-0.35cm]
    	&
    \multicolumn{1}{c}{$E_\nonrel$}	&	
    \multicolumn{1}{c}{$-\frac{1}{8}\sum_{i} \langle (\bos{p}_i^2)^2 \rangle_\nonrel$}	&	
    \multicolumn{1}{c}{$\sum_{i,a} Z_a \langle \delta(\bos{r}_{ia})\rangle_\nonrel$}	&	
    \multicolumn{1}{c}{$\langle \delta(\bos{r}_{12})\rangle_\nonrel$} &
    \multicolumn{1}{c}{$\langle H_\text{OO}\rangle_\nonrel$} \\
    \hline\\[-0.35cm]
    H$^-$	    & $-$0.527 751 016	&	$-$0.615 429	&	0.328 980	&	0.002 742	&	$-$0.008 875	\\
    He ($2S$)     	& $-$2.145 974 046	&	$-$10.273 526	&	5.234 159	&	0.008 659	&	$-$0.009 253	\\    
    He ($1S$)  & $-$2.903 724 377	&	$-$13.517 694	&	7.238 965	&	0.106 445	&	$-$0.139 095	\\
    Li$^+$	    & $-$7.279 913 410	&	$-$77.562 651	&	41.067 671	&	0.534 083	&	$-$0.427 992	\\
    Be$^{2+}$	& $-$13.655 566 234	&	$-$261.660 823	&	137.491 296	&	1.523 969	&	$-$0.878 771	\\
    \hline \\[-0.35cm]													
    H$_2$ $^\text{a}$ 	& $-$1.174 475 714	&	$-$1.653 578	&	0.918 653	&	0.016 768	&	$-$0.047 635	\\
    H$_3^+$ $^\text{a}$ 	& $-$1.343 835 625	&	$-$1.932 048	&	1.088 845	&	0.018 358	&	$-$0.057 218	\\
    HeH$^+$ $^\text{a}$ 	& $-$2.978 706 599	&	$-$13.406 476	&	7.208 549	&	0.101 223	&	$-$0.141 242	\\
    \hline\hline\\[-0.35cm]
    	&	
    \multicolumn{1}{c}{$\langle H_\DC^{(2)} \rangle_\nonrel$} & 
    \multicolumn{1}{c}{$\langle H_\text{B}^{(2)} \rangle_\nonrel$} & 
    \multicolumn{1}{c}{$\langle H_\DC^{(2)}+H_\text{B}^{(2)} \rangle_\nonrel$} & 	
    \multicolumn{1}{c}{$E^{(2)}_\DC$} &	
    \multicolumn{1}{c}{$E^{(2)}_\DCB$}	\\
    \hline\\[-0.35cm]
    H$^-$ & $-$0.107 283	&	0.008 355	&	$-$0.098 928	&	$-$0.527 756 729	&	$-$0.527 756 284	\\
    He ($2S$)     	& $-$2.078 929	&	0.045 150	&	$-$2.033 779	&	$-$2.146 084 751	&	$-$2.146 082 347	\\    
    He ($1S$) & $-$2.481 162	&	0.529 721	&	$-$1.951 442	&	$-$2.903 856 502	&	$-$2.903 828 294	\\
    Li$^+$ & $-$14.731 576	&	2.927 751	&	$-$11.803 825	&	$-$7.280 697 887	&	$-$7.280 541 980	\\
    Be$^{2+}$ & $-$50.477 690	&	8.696 608	&	$-$41.781 082	&	$-$13.658 254 239	&	$-$13.657 791 133	\\
    \hline\\[-0.35cm]
    H$_2$~$^\text{a}$  &	$-$0.263 240	&	0.057 725	&	$-$0.205 516	&	$-$1.174 489 732	&	$-$1.174 486 658	\\
    H$_3^+$~$^\text{a}$  & $-$0.279 367	&	0.058 126	&	$-$0.221 240	&	$-$1.343 850 502	&	$-$1.343 847 407	\\ 
    HeH$^+$~$^\text{a}$  & $-$2.401 316	&	0.494 763	&	$-$1.906 553	&	$-$2.978 834 472	&	$-$2.978 808 125	\\
    \hline\hline\\[-0.35cm]
  \end{tabular}
}
  \begin{flushleft}
    {\footnotesize  
    \vspace{-0.20cm}  
    $^\text{a}$ $R_\text{pp}=1.4$~bohr, $R_\text{pp}=1.65$~bohr, $R_\text{pp}=1.46$~bohr.}
  \end{flushleft}  
\end{table}

\clearpage
%\bibliography{references}

\end{document}